\documentclass[aps,twocolumn,superscriptaddress]{revtex4-1}
\usepackage{graphicx}
\usepackage{amsmath}
\usepackage{natbib}
\usepackage{times}
\usepackage[colorlinks,citecolor=blue]{hyperref}
\usepackage[left=2cm,right=1cm,bottom=2cm,top=2cm]{geometry}
\usepackage[normalem]{ulem} 

\begin{document}

\title{Path Integral Monte Carlo and Density Functional Molecular Dynamics Simulations of Warm, Dense MgSiO$_3$ }

\author{Felipe Gonz\'alez-Cataldo}
\email{f\_gonzalez@berkeley.edu}
\affiliation{Department of Earth and Planetary Science, University of California, Berkeley, California 94720, USA}
\author{Fran\c{c}ois Soubiran}
\affiliation{Laboratoire de g\'eologie de Lyon, CNRS UMR 5276, Ecole Normale Sup\'erieure de Lyon, Universit\'e Claude Bernard Lyon 1, 46 All\'ee d'Italie, 69364 Lyon Cedex 07, France}
\affiliation{Department of Earth and Planetary Science, University of California, Berkeley, California 94720, USA}
\author{Henry Peterson}
\affiliation{Department of Earth and Planetary Science, University of California, Berkeley, California 94720, USA}
\author{Burkhard Militzer}
\affiliation{Department of Earth and Planetary Science, University of California, Berkeley, California 94720, USA}
\affiliation{Department of Astronomy, University of California, Berkeley, California, USA}

\date{\today}

\begin{abstract} 
  In order to provide a comprehensive theoretical description of
  MgSiO$_3$ at extreme conditions, we combine results from path
  integral Monte Carlo (PIMC) and density functional molecular
  dynamics simulations (DFT-MD) and generate a consistent equation of state for
  this material. We consider a wide range of temperature and density
  conditions from 10$^4$ to 10$^8$ K and from 0.321 to 64.2
  g$\,$cm$^{-3}$ (0.1- to 20-fold the ambient density). We study how
  the L and K shell electrons are ionized with increasing temperature
  and pressure.  We derive the shock Hugoniot curve and compare with
  experimental results. Our Hugoniot curve is in good agreement
	with the experiments, and we predict a broad compression maximum 
  that is dominated by the K shell ionization of all
  three nuclei while the peak compression ratio of 4.70 is obtained
  when the Si and Mg nuclei are ionized. Finally we analyze the heat
  capacity and structural properties of the liquid.
\end{abstract}


\maketitle

\section{Introduction}
The equation of state (EOS) of materials in the regime of warm dense matter
is fundamental to model planetary interiors~\cite{Mazevet2018,Seager2007,Benuzzi2014},
astrophysical processes~\cite{Cotelo2011,Chabrier2002},
interpret shock-wave experiments~\cite{Kirsch2019,Millot2015},
and understand the physics of inertial confinement fusion experiments~\cite{Zhang2018,Betti2016,Seidl2009,Miyanishi2015}.
Novel experiments and computational techniques have allowed the study of the
properties of matter at extreme conditions and produce EOS of materials in a wide
range of temperatures and densities. 
Among the computational techniques are path integral Monte Carlo (PIMC)
simulations~\cite{Feynman1979,Ce91,PC94,Rickwardt2001,MC00},
which have provided a unique insight into the properties of matter
at extreme temperature and pressure conditions relevant to 
fusion experiments, where a detailed description of dense plasmas
is required to understand the underlying physics.
There has been a considerable effort to study the properties of materials
heavier than hydrogen and helium in the warm dense matter regime with these
techniques, materials such as lithium fluoride~\cite{Driver2017},
boron~\cite{Zhang2018}, aluminum~\cite{Driver2018},
hydrocarbons~\cite{Driver2017b}, and superionic
water~\cite{Wilson2013,Millot2018}.  However, the properties of triatomic
materials, such as MgSiO$_3$, have not been studied.

Enstatite (MgSiO$_3$) is key material for planetary science and shock
physics~\cite{Duffy2019,Duffy2015,Luo2004,Millot2015,Kraus2012}.  It is one of
the few silicate minerals that has been observed in crystalline form outside
the Solar System~\cite{Molster2001}, and is assumed to be one fundamental
building block in planetary formation~\cite{Valencia2010,Bolis2016}.  Along
with forsterite (Mg$_2$SiO$_4$), it is one of the most abundant materials in
the Earth's mantle, and it is also expected to be present in super-Earth
planets~\cite{Molster2001,Valencia2009}.  The properties of silicates at
conditions existing at planetary interiors are poorly known because reaching
Mbar pressures and 5000--10000 K temperatures in the laboratory presents a
serious challenge.

Recent ramp compression experiments at the National Ignition Facility (NIF) and
the \emph{Ligne d'Int\'egration Laser} (LIL) facility have explored the
properties of silicates and iron at the conditions encountered in planetary
interiors, studying their metallization and dissociation up to 15
Mbar~\cite{Benuzzi2014,Smith2018,Hicks2006}.  Under ramp compression, the
system follows a thermodynamic path that is
quasi-isentropic~\cite{Swift2008,Rothman2005,Bradley2009,Smith2014,Smith2018},
as the heat generated is significantly lower than in shock compression.  This
is ideal to reach the pressures and temperatures present at the interior of
super-Earth planets~\cite{Seager2007,Wagner2012}.  How close the compression
path follows an isentrope depends on the sample properties and on the details
of the experiments.  It is therefore beneficial to compute the isentropes with
first-principle computer simulations in order to guide interpretations of the
experimental findings. Isentropes are also of fundamental importance in
planetary science because planets cool convectively, thus, most of the
interiors are assumed to be
adiabatic~\cite{hubbard_planets,Valencia2009,Seager2007,Wagner2012}. However,
there are known exceptions, such as the boundary layer in the Earth mantle,
where temperature rises superadiabatically because of the high mantle viscosity,
and the outermost atmospheres of giant planets, where heat is carried
radiatively.

In recent laser-shock experiments, the equations of state of enstatite
and forsterite on the principal Hugoniot curve have been measured up
to 950 GPa and 30000 K~\cite{Spaulding2012,Root2018,Bolis2016},
suggesting a metallic-like behavior in liquid MgSiO$_3$ over a large
pressure-temperature regime.  Finding signatures of melting along the
Hugoniot curves of silicates is fundamental to understand the dynamics
of the Earth's lower mantle~\cite{Hicks2006,Mosenfelder2009}, as well
as understanding the rich phase diagram of MgSiO$_3$, which undergoes
a series of phase transitions before partitioning into Mg and
SiO$_2$~\cite{Tsuchiya2004,Umemoto2017}.  However, the behavior of
this mineral at temperatures relevant to the conditions of shock
experiments where ionization of the electronic shells take place, is
unknown.

Recent \emph{ab initio} simulations have shown that liquid silicates can exhibit
very high conductivity at high pressure, which implies that super-Earths can
generate magnetic fields in their mantle~\cite{Soubiran2018}.  Therefore, it is
desirable to have a first-principles EOS derived for much higher temperature and
density conditions that span regimes of condensed matter, warm dense matter
(WDM), and plasma physics in order to be used as a reference for shock
experiments and hydrodynamic simulations.  In recent works, a first-principles
framework has been developed to compute consistent EOS across a wide range of
density-temperature regimes relevant to WDM by combining results from
state-of-the-art path integral Monte Carlo (PIMC) and DFT-MD simulation methods
for first- \cite{Driver2012} and second-row \cite{MilitzerDriver2015,Soubiran2019} elements.

In this paper, we apply our PIMC and DFT-MD methods to compute the EOS
and plasma properties of MgSiO$_3$ across a wide density-temperature
range. We study the evolution of the plasma structure and ionization
over the WDM regime.  Finally, we compare our PIMC/DFT-MD shock
Hugoniot curves with widely used models and experiments.

\section{Methods}


We perform first-principles computer simulations of warm-dense
MgSiO$_3$ using two different simulation methods: path integral Monte
Carlo (PIMC) and Kohn-Sham density functional theory molecular
dynamics (DFT-MD) simulations.  PIMC is a state-of-the-art
first-principles technique for computing the properties of interacting
quantum systems at finite temperature. The basic techniques for
simulating bosonic systems were developed in Ref. \cite{PC84} and
reviewed in Ref. \cite{Ce95}. Subsequently the algorithm was
generalized to fermion systems using the {\em restricted} path
integral method.  The first results of this simulation method were
reported in the seminal work on liquid $^3$He~\cite{Ce92} and dense
hydrogen~\cite{PC94}. A review of the algorithm is given in
Ref.~\cite{Ce96}. In subsequent articles, this method was applied to
study hydrogen~\cite{Ma96,Mi99,MC00,MC01,Mi01},
helium~\cite{Mi06,Mi09,Mi09b}, hydrogen-helium mixtures~\cite{Mi05}
and one-component plasmas~\cite{JC96,MP04,MP05}. In recent years, the
method was extended to simulate plasmas of various first-row
elements~\cite{Benedict2014,DriverNitrogen2016,Driver2017,ZhangCH2017,ZhangCH2018,Zhang2018}
and with the development of Hartree-Fock nodes, the simulations of
second-row elements became
possible~\cite{MilitzerDriver2015,Hu2016,ZhangSodium2017,Driver2018}.

This method is based on the thermal density matrix of a quantum
system, $\hat\rho=e^{-\beta \hat{\cal H}}$, that is expressed as a
product of higher-temperature matrices by means of the identity
$e^{-\beta \hat{\cal H}}=(e^{-\tau \hat{\cal H}})^M$, where
$\tau\equiv\beta/M$ represents the time step of a path integral in
imaginary time. The path integral emerges when the operator $\hat\rho$
is evaluated in real space,
\begin{equation}
\left<\mathbf R|\hat\rho| \mathbf R'\right>=\frac{1}{N!}\sum_{\mathcal P}(-1)^{\mathcal P}\oint_{\mathbf R\to\mathcal P\mathbf R'}\mathbf{dR}_t\, e^{-S[\mathbf R_t]}.
\label{PI}
\end{equation}
Here, we have already summed over all permutations, $\mathcal P$, of
all $N$ identical fermions in order project out all antisymmetric
states.  For sufficiently small time steps, $\tau$, all many-body
correlation effects vanish and the action, $S[\mathbf R_t]$, can be
computed by solving a series of two-particle
problems~\cite{PC84,Na95,BM2016}. The advantage of this path integral
approach is that all many-body quantum correlations are recovered
through the integration over all paths. The integration also enables
one to compute quantum mechanical expectation values of themodynamic
observables, such as the kinetic and potential energies, pressure,
pair correlation functions and the momentum
distribution~\cite{Ce95,Militzer2019}.  Most practical implementations of
the path integral techniques rely on Monte Carlo sampling techniques
due to the high dimensionality of the integral and, in
addition, one needs to sum over all permutations. The method becomes
increasingly efficient at high temperature because the path the length
of the paths scales like $1/T$. In the limit of low temperature, where
few electronic excitations are present, the PIMC method becomes
computationally demanding and the MC sampling can become inefficient.
However, the PIMC method avoids any exchange-correlation approximation
and the calculation of single-particle eigenstates, which are deeply
embedded in all Kohn-Sham DFT calculations. 

The only uncontrolled approximation within fermionic PIMC calculations
is the use of the fixed-node approximation, which restricts the paths
in order to avoid the well-known fermion sign
problem~~\cite{Ce91,Ce92,Ce96}. Addressing this problem in PIMC is
crucial, as it causes large fluctuations in computed averages due to
the cancellation of positive and negative permutations in
Eq.~\eqref{PI}. We solve the sign problem approximately by restricting
the paths to stay within our Hartree-Fock
nodes~\cite{MilitzerDriver2015,ZhangCH2017,Driver2018}.  We enforced the nodal
constraint in small imaginary time steps of $\tau=1/8192$ Ha, while
the pair density matrices were evaluated in steps of 1/1024 Ha.  This
results in using between 1200 and 12 time slices for the temperature
range that studied with PIMC simulations here. These choices converged
the internal energy per atom to better than 1\%.  We have shown the
associated error is small for relevant systems at sufficiently high
temperatures~\cite{Driver2012,Ce91,Ce96}.

On the other hand, Kohn-Sham
DFT-MD~\cite{Hohenberg1964,Kohn1965,Mermin1965} is a well-established
theory that has been widely applied to compute the EOS of condensed
matter as well as warm and hot, dense
plasmas~\cite{Root2010,Wang2010,Mattsson2014,Zhang2018}.  It is a
suitable option to derive the EOS because it accounts for both the
electronic shells and bonding effects.  The main approximation in
DFT-MD is the use of an approximate exchange-correlation (XC)
functional.  Although at temperatures relevant to WDM, the error in
the XC functional is small relative to the total energy, which is the
most relevant quantity for the EOS and derivation of the shock
Hugoniot curve~\cite{Karasiev2016}.

Still, standard Kohn-Sham DFT-MD simulations become computationally
inefficient at high temperatures ($T>10^6$ K) because it requires one
to explicitly compute all fully and partially occupied electronic
orbitals, which becomes increasingly demanding as temperature
increases.  The number of occupied bands increases unfavorably with
temperature, scaling approximately as $\sim T^{3/2}$.  Accuracy is
also compromised at high temperatures.  The excitation of the inner
electrons, which are typically frozen by the pseudopotentials, may
contribute to the pressure and energy of the system as inner
electronic shells become partially ionized with increasing
temperature.  In contrast, PIMC is an all-electrons method that
increases in efficiency with temperature (scaling as $1/T$) as quantum
paths become shorter and more classical in nature.

Consequently, our approach consist in performing simulations along different isochores of MgSiO$_3$,
using PIMC at high temperatures ($1.3\times10^6$ K $\leq T\leq5.2\times10^8$ K ) and DFT-MD at
low temperatures ($1.0\times10^4\text{ K }\leq T \leq 1.0\times10^6$ K). We show the two methods produce consistent
results at overlapping temperature regimes.


For PIMC simulations, we use the CUPID code~\cite{MilitzerThesis} with
Hartree-Fock nodes.  For DFT-MD simulations, we employ Kohn-Sham DFT simulation
techniques as implemented in the Vienna Ab initio Simulation Package
(VASP)~\cite{VASP} using the projector augmented-wave (PAW)
method~\cite{PAW,Kresse1999}, and molecular dynamics is performed in the NVT
ensemble, regulated with a Nos\'e thermostat.  Exchange-correlation effects are
described using the Perdew, Burke, and Ernzerhof~\cite{PBE} (PBE) generalized
gradient approximation (GGA).  The pseudopotentials used in our DFT-MD
calculations freeze the electrons of the 1s orbital, which leaves 10, 12, and
6 valence electrons for Mg, Si, and O atoms, respectively.  Electronic wave
functions are expanded in a plane-wave basis with a energy cut-off as high as
7000 eV in order to converge total energy. Size convergence tests with up to a
65-atom simulation cell at temperatures of 10\,000 K and above indicate that
pressures are converged to better than 0.6\%, while internal energies are
converged to better than 0.1\%.  We find, at temperatures above 500\,000 K,
that 15-atom supercells are sufficient to obtain converged results for both
energy and pressure, since the kinetic energy far outweighs the interaction
energy at such high temperatures~\cite{Driver2018,Driver2015}. The number of
bands in each calculation was selected such that orbitals with occupation as
low as $10^{-4}$ were included, which requires up to 14\,000 bands in an
15-atom cell at $2\times10^6$ K and two-fold compression.  All simulations are
performed at the $\Gamma$ point of the Brillouin zone, which is sufficient for
high temperature fluids, converging total energy to better than 0.01\% compared
to a grid of $k$-points.

\section{Equation of State Results}\label{sec:EOS}

\begin{figure}[!hbt]
\includegraphics[width=9cm]{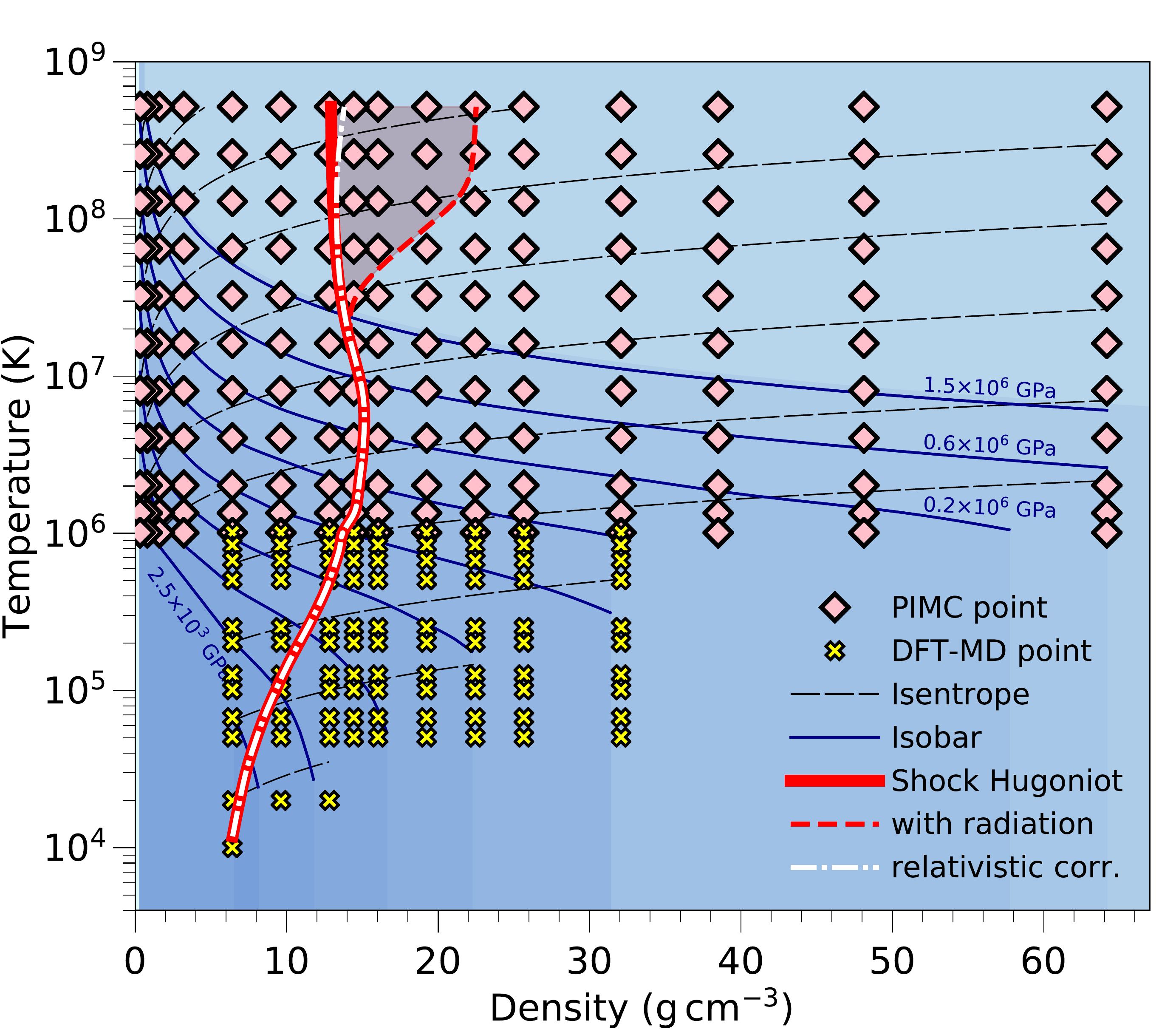}
\caption{
Temperature-density conditions of our DFT-MD and PIMC simulations along with
computed isobars, isentropes and three principal shock Hugoniot curve that were
derived for an initial density of
$\rho_0=3.207911$ g cm$^{-3}$ ($V_0=51.965073$~\AA/f.u.).
\label{fig:Tvsrho}
}
\end{figure}

\begin{figure}[!hbt] 
  \includegraphics[width=9cm]{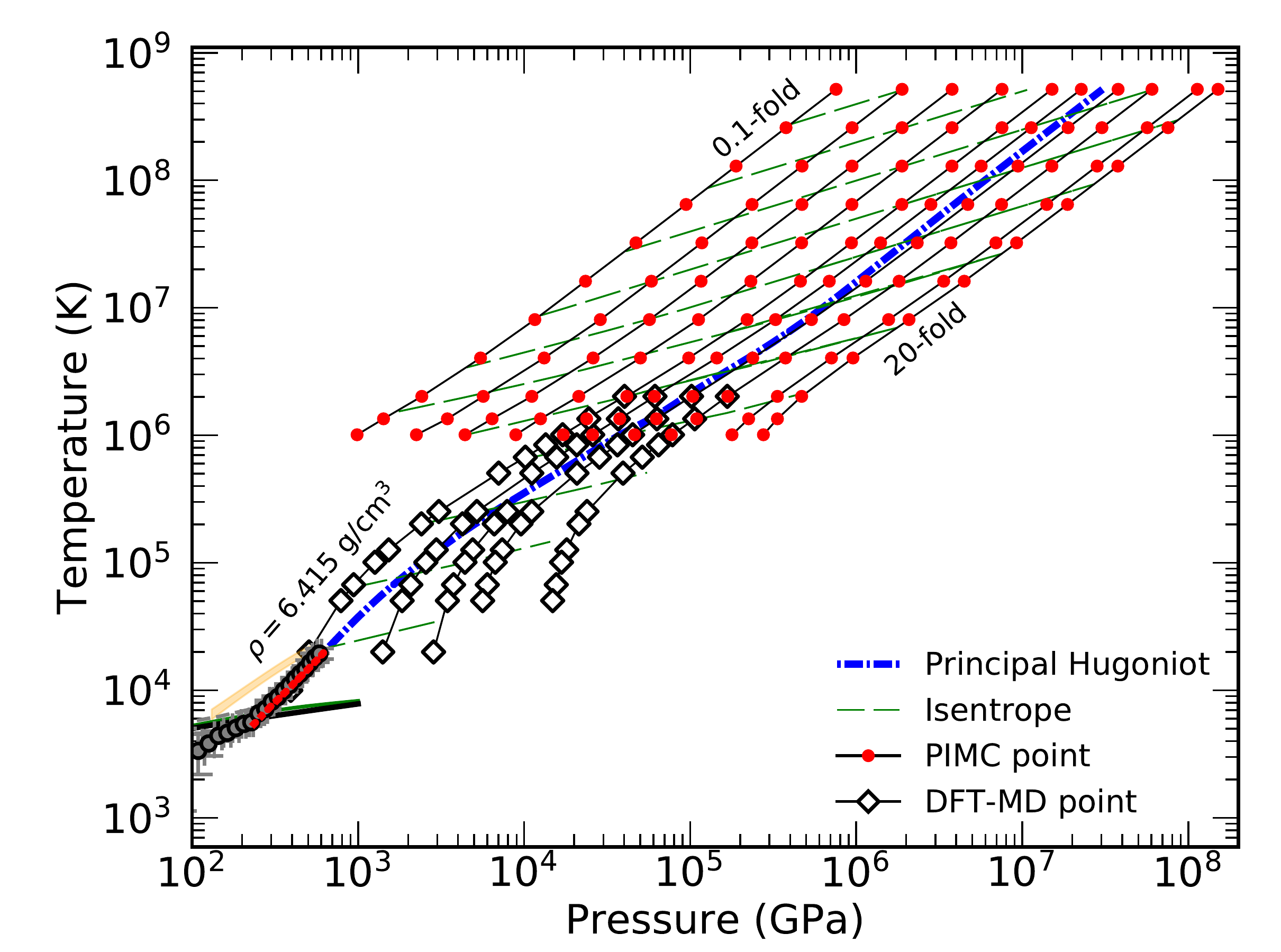}
  \includegraphics[width=9cm]{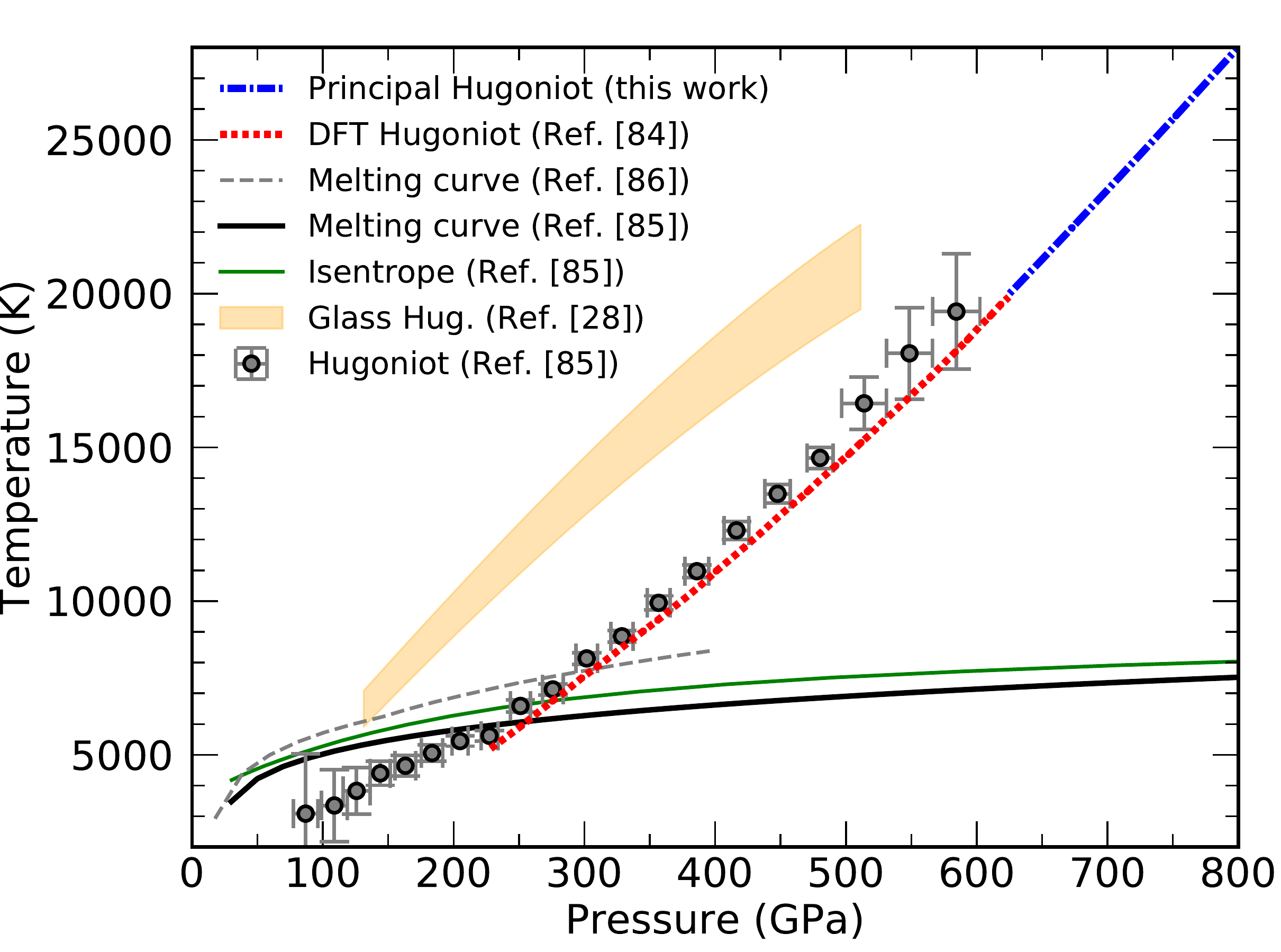}
  \caption{ Temperature-pressure conditions for the PIMC and DFT-MD
    calculations along isochores corresponding to the densities of
    0.1-fold (0.321 g cm$^{-3}$) to 20-fold (64.20 g cm$^{-3}$). The
    blue dash-dotted line shows the principal Hugoniot curve of
    MgSiO$_3$ obtained from our simulations, using an initial density
    of $\rho_0=3.207911$ g cm$^{-3}$ ($V_0=51.965073$~\AA/f.u.).
    The red dashed line corresponds to the Hugoniot curve from
    Ref.~\cite{Militzer2013d}, calculated from DFT-MD simulations.
    Experimental measurement of the
    principal Hugoniot curve from Ref.~\cite{Fratanduono2018},
    an isentrope derived from this experiment
    (solid green line), and the Hugoniot curve for MgSiO$_3$
    glass~\cite{Bolis2016} (orange region) are shown for
    reference. The melting line of MgSiO$_3$ derived
    from two-phase simulations~\cite{Belonoshko2005} is shown
    in dashed grey line, while the melting curve derived from
    shock experiments~\cite{Fratanduono2018} is represented
    by the thick black line.
  }\label{fig:PvsT} 
\end{figure}

In this section, we combine results from our PIMC and DFT-MD
simulations in order to provide a consistent EOS table spanning the
warm dense matter and plasma regimes. Computations were performed for a
series of densities and temperatures ranging from 0.321--64.16
g$\,$cm$^{-3}$ and $10^4$--$10^{8}$ K.
The full range of our EOS data points is shown in temperature-density
space in Fig.~\ref{fig:Tvsrho} and in temperature-pressure space in
Fig.~\ref{fig:PvsT}.

In order to put the VASP PBE pseudopotential energies on the same
scale as the all-electron PIMC calculations, we shifted all VASP
DFT-MD energies by $\Delta E=-$713.777558 Ha/atom. This shift was derived by
performing all-electron calculations for the isolated
non-spin-polarized Mg, Si, and O atoms with the OPIUM
code~\cite{OPIUM} and comparing the results with corresponding VASP
calculations.

\begin{figure}[!hbt]
  \includegraphics[width=9cm]{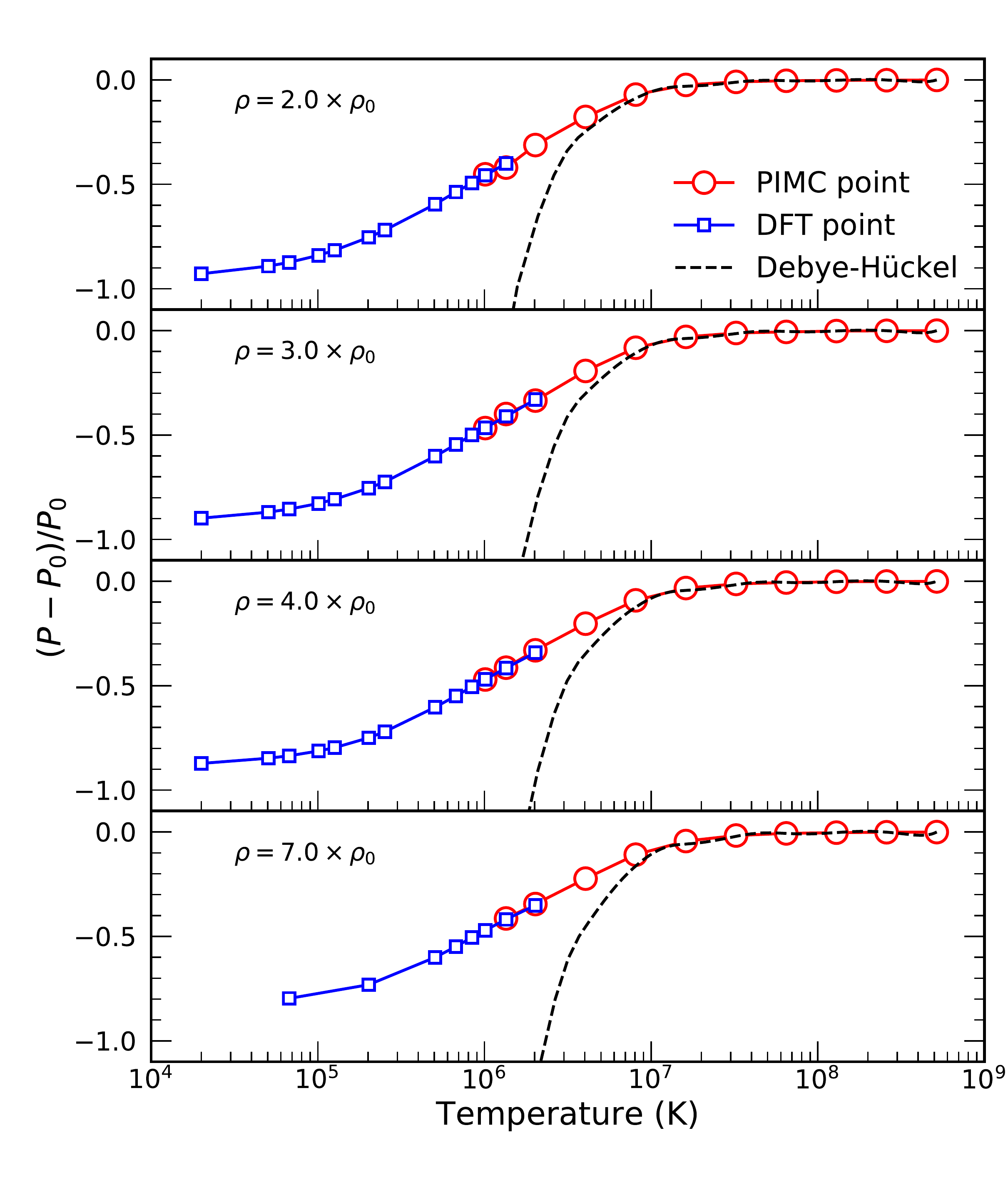} \caption{
    Excess pressure as a function of temperature relative to the ideal
    Fermi gas, computed with PIMC, DFT-MD, and the Debye-H\"uckel
    plasma model. The results are plotted for densities of (a) 6.4,
    (b) 3.651, (c) 7.582, and (d) 15.701 g cm$^{-3}$.\label{PRelvsT}}
\end{figure}

\begin{figure}[!hbt]
  \includegraphics[width=9cm]{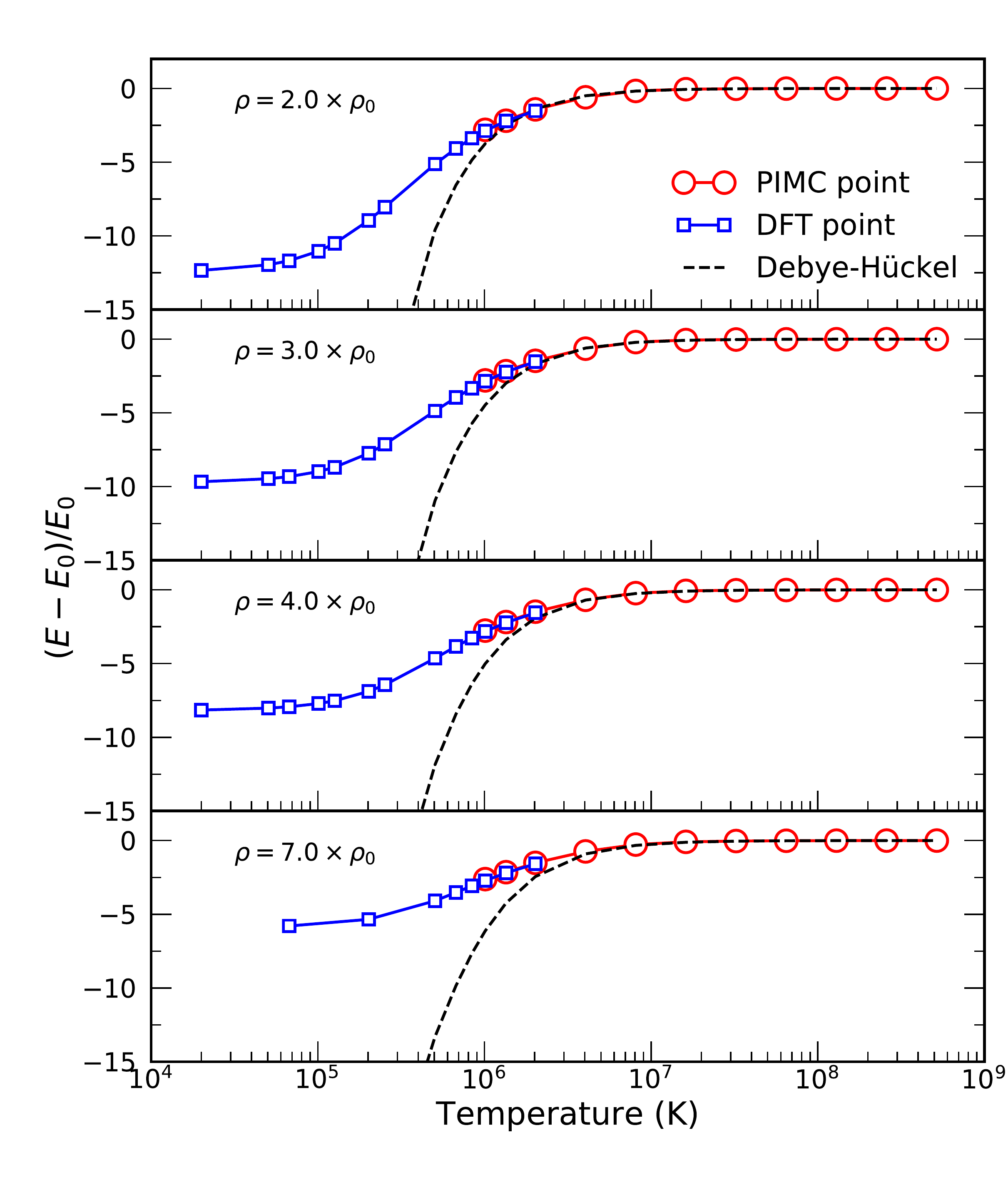} \caption{
    Excess internal energy, relative to the ideal Fermi gas, computed
    with PIMC, DFT-MD, and the Debye-H\"uckel plasma model. As in the
    corresponding Fig.~\ref{PRelvsT}, the results are plotted for
    densities of (a) 6.4, (b) 3.651, (c) 7.582, and (d) 15.701 g
    cm$^{-3}$ as a function of temperature.}\label{ERelvsT}
\end{figure}

In order to analyze the consistency of our EOS data sets,
Figs.~\ref{PRelvsT} and~\ref{ERelvsT} display the
pressure and internal energy, respectively, along three isochores from PIMC, DFT-MD,
and the classical Debye-H\"uckel plasma model~\cite{Debye1923} as a
function of temperature. The pressures, $P$, and internal energies,
$E$, are plotted relative to a fully ionized Fermi gas of electrons
and ions with pressure $P_0$ and internal energy $E_0$ in order to
compare only the excess contributions that are the result from the
particle interactions.

With increasing temperature, these contributions gradually decrease
from the strongly interacting condensed matter regime, where chemical
bonds and bound states dominate, to the weakly interacting, fully
ionized plasma regime. There, the PIMC results converge to predictions
from the classical Debye-H\"uckel model. As expected, the
Debye-H\"uckel model becomes inadequate for lower temperatures
($T < 10^7$ K) since it fails to treat bound electronic states. While
the temperature range over which PIMC EOS data are needed to fill the
gap between DFT-MD and Debye-H\"uckel model (approximately
from $2 \times 10^6$ to $1 \times 10^7$ K) is relatively small compared to
the entire temperature range under consideration, this temperature
interval encompasses significant portions of K shell ionization
regime, which is precisely where the full rigor of PIMC simulations
are needed to acquire an accurate EOS table.

Figs.~\ref{PRelvsT} and \ref{ERelvsT} show a consistent EOS over a wide
density-temperature range, where PIMC and DFT-MD simulations provide consistent results
in the overlapping range of 1--2 $\times 10^6$ K.
At these temperatures, the pressures predicted by PIMC and DFT-MD differ by less
than 3\%, with the exception of $10^6$ K at 7-fold compression,
where we obtained a difference of 5.3\% in the pressure.
We attribute this difference to the known loss of accuracy of PIMC at low temperature.
However, we do not observe this large difference at any other density.

The total energies predicted by DFT-MD are also in good agreement with those
predicted by PIMC (see Fig.~\ref{ERelvsT}), with differences generally between
1.5--6.5 Ha/atom (3--6\%). Larger energy differences are observed at
$2\times10^6$ K, where DFT-MD seems unable to reproduce the energies predicted
by PIMC, within the error bars.
At this temperature, we observe a systematic energy offset of
6.5--8.5 Ha/atom (11--23\%) as the density increases.
These errors are mostly due to the use of pseudopotentials used in DFT
simulations, where inner electrons are bound to the nucleus and cannot be
excited to contribute to the energy, resulting in an underestimation of the
total energy of the system.  We will come back to this point when we discuss
the ionization of the electronic shells (Fig.~\ref{fig:N(r)}) in the next
section.

\begin{figure}[!hbt]
\includegraphics[width=9cm]{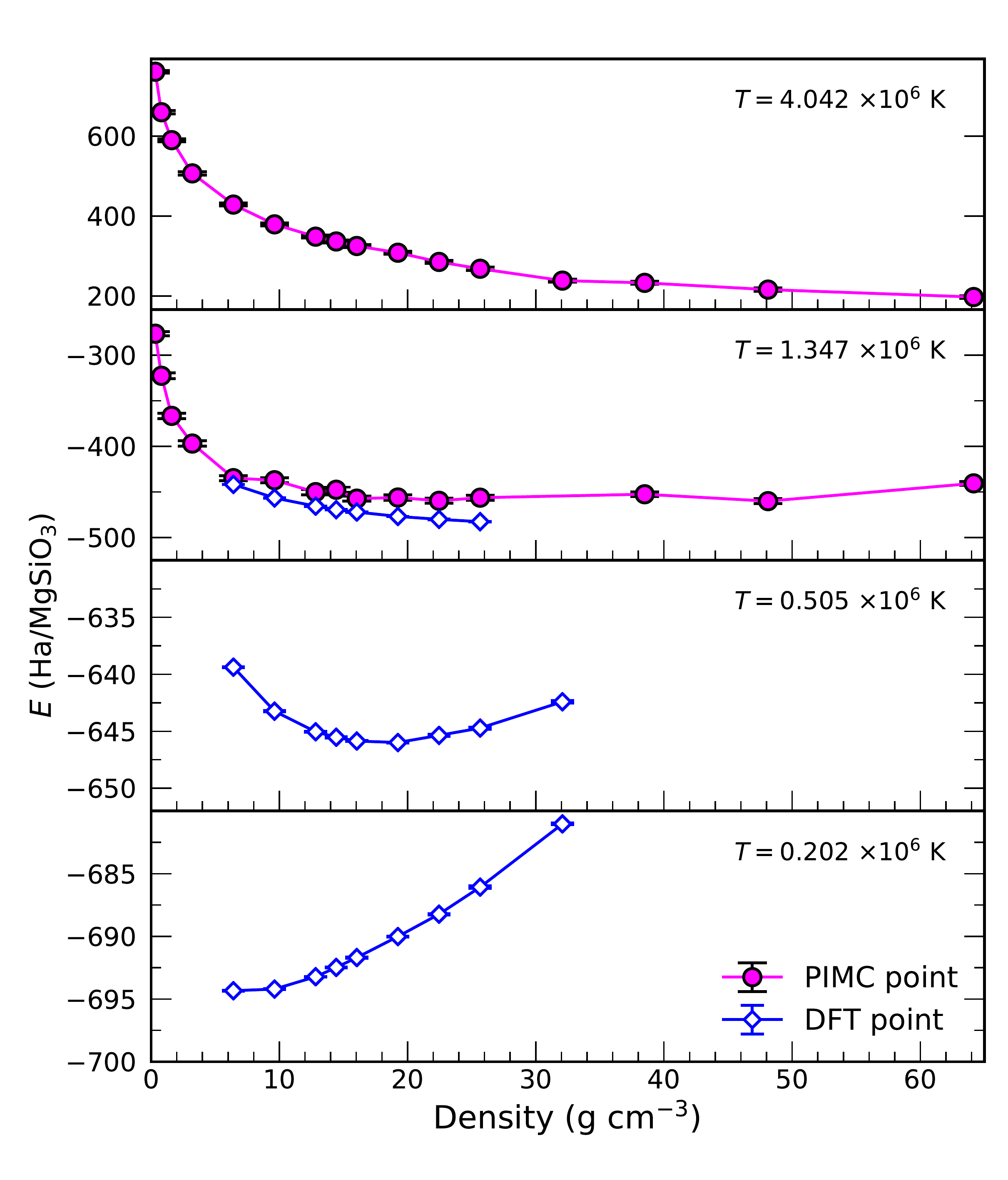}
\caption{Total internal energy as a function of density, computed with PIMC and DFT-MD. }\label{Ediffvsrho}
\end{figure}
\begin{figure}[!hbt]
\includegraphics[width=9cm]{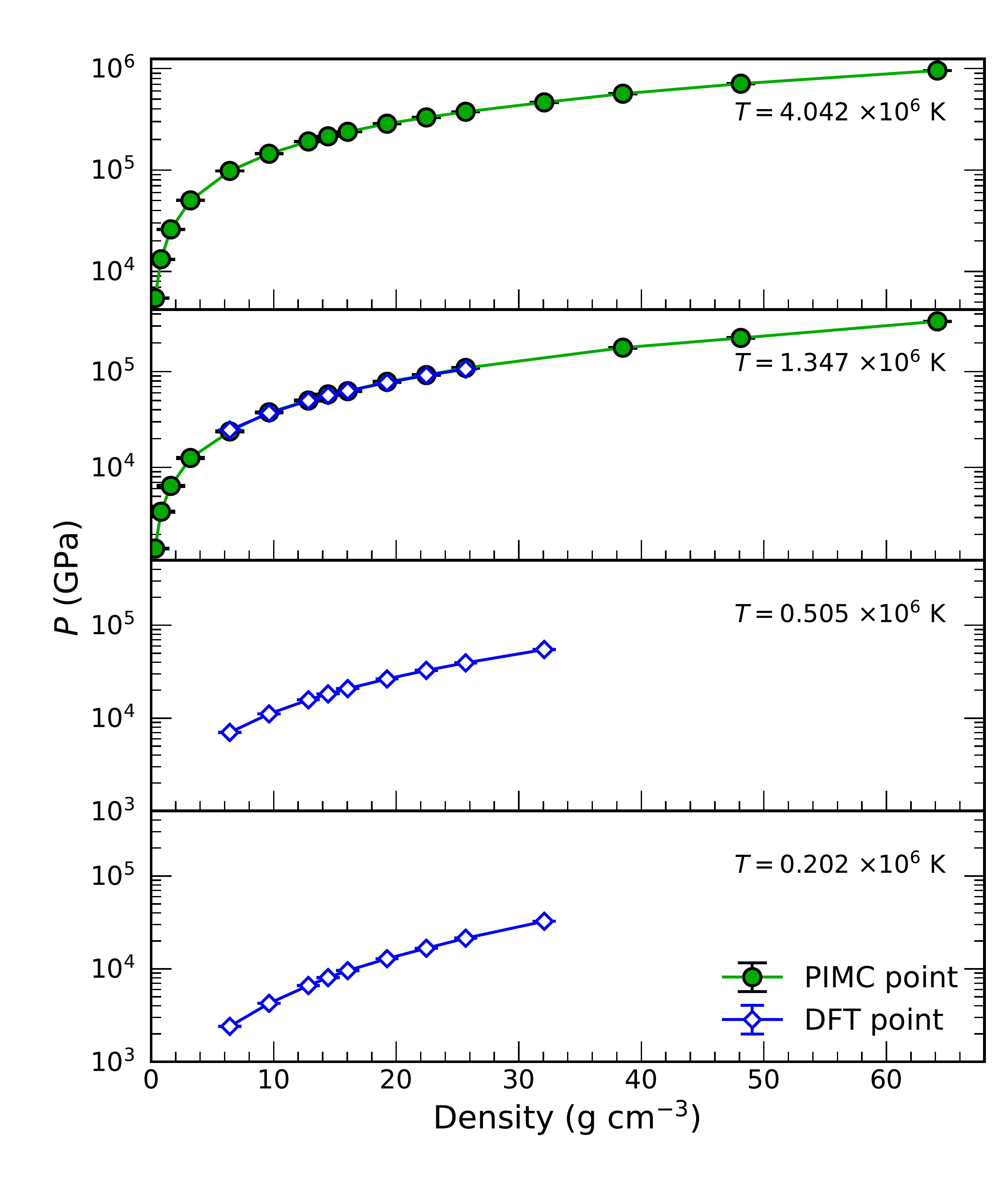}
\caption{Pressure as a function of density, computed with PIMC and DFT-MD.}\label{Pdiffvsrho}
\end{figure}

In Fig.~\ref{Ediffvsrho} and ~\ref{Pdiffvsrho}, we show
  the total energy and pressure as a function of density for a number
  of temperatures.  While pressure increases with density, we
    find that all the $E(\rho)_T$ curves have a minimum. With
    increasing temperature, the location of this minimum shifts
    towards high densities.  At low density, the slope
    $\left(\frac{\partial E}{\partial \rho}\right)_T$ is negative
    because the system is more ionized, as we will
    discuss in section~\ref{sec:Ionization}. At high density, the
    slope $\left(\frac{\partial E}{\partial \rho}\right)_T$ is
    positive for two possible reasons. First there is the confinement
    effect, which increases the kinetic energy of the free electrons
    and, second, the orbitals of the bound electrons hybridize and may
    even be pushed into the continuum of free electronic states, which
    is commonly referred to as pressure ionization.

Using Maxwell relations, one can infer that this 
    energy minimum corresponds to the point where the thermal
    pressure coefficient,
    $\beta_V\equiv\left.\frac{\partial P}{\partial T}\right|_V$, is
    equal to the ratio between pressure and temperature, because
\begin{equation}\label{eq:CondEmin}
\left(\frac{\partial E}{\partial V}\right)_T = T\left[\left(\frac{\partial P}{\partial T}\right)_V -\frac PT\right]= T\left[\beta_V-\frac PT\right] =0.
\end{equation}
This derivative vanishes if
\begin{equation}\label{eq:lnPlnT=1}
\left(\frac{\partial \ln P}{\partial \ln T}\right)_V=1 \;.
\end{equation}

This condition is trivially fulfilled for an ideal gas, that
satisfies $\left.\frac{\partial E}{\partial V}\right|_T=0$
everywhere. At very high temperature, where ionization is complete, we
find that MgSiO$_3$ starts behaving similar to an ideal gas and the
isochores, that we show in $\ln T$-$\ln P$ space in
Fig.~\ref{fig:PvsT}, have a slope of approximately 1.
%
When Eq.~\ref{eq:lnPlnT=1} is satified, we obtain a minimum in the
$E(\rho)_T$ curve.
For example, at $T=0.202\times10^6$, we find an energy minimum in
Fig.~\ref{Ediffvsrho} around $\rho\approx 6.42$ g cm$^{-3}$ while
$\left.\frac{\partial \ln P}{\partial \ln T}\right|_V$ becomes 1 in
Fig.~\ref{fig:PvsT} for the same conditions.

%

We note that the overall agreement between PIMC and DFT-MD provides validation
for the use of zero-temperature exchange correlation functionals in warm dense
matter applications and the use of the fixed-node approximation in PIMC in the
relevant temperature range. At temperatures lower than the overlapping regime,
PIMC results become inconsistent with DFT-MD results because the nodal
approximation in PIMC simulations is no longer appropriate.  Nevertheless, the
validity of our EOS is not affected by these discrepancies, as we are able to
build a consistent interpolation that spans across all temperatures. 

The isochoric Gr\"uneisen parameter,
\begin{equation}\label{eq:Gruneisen}
\gamma=V\left(\frac{\partial P}{\partial E}\right)_V
=\frac{V}{C_V}\left(\frac{\partial P}{\partial T}\right)_V
=-\left(\frac{\partial \ln T}{\partial \ln V}\right)_S,
\end{equation}
is a useful quantity to model material properties, since it usually does not
significantly depend on temperature.  It is the key parameter of the
Mie-Gr\"uneisen model, which is often used in shock experiments to model the
EOS of solids and
liquids~\cite{Mcwilliams2012,Mosenfelder2009,Fratanduono2018,Spaulding2012} and
obtain related properties, such as the specific heat, melting temperature and, in
general, to infer how pressure depends on temperature along different
thermodynamic paths.
The Gr\"uneisen parameter can also be inferred from the shock Hugoniot
curve~\cite{Fratanduono2018} and, by means of Eq.~\eqref{eq:Gruneisen}, can
be used to obtain isentropic paths, such as the temperature profile in magma
oceans and ramp compression curves~\cite{Rothman2005,Smith2018}.

\begin{figure}[!hbt]
\includegraphics[width=9cm]{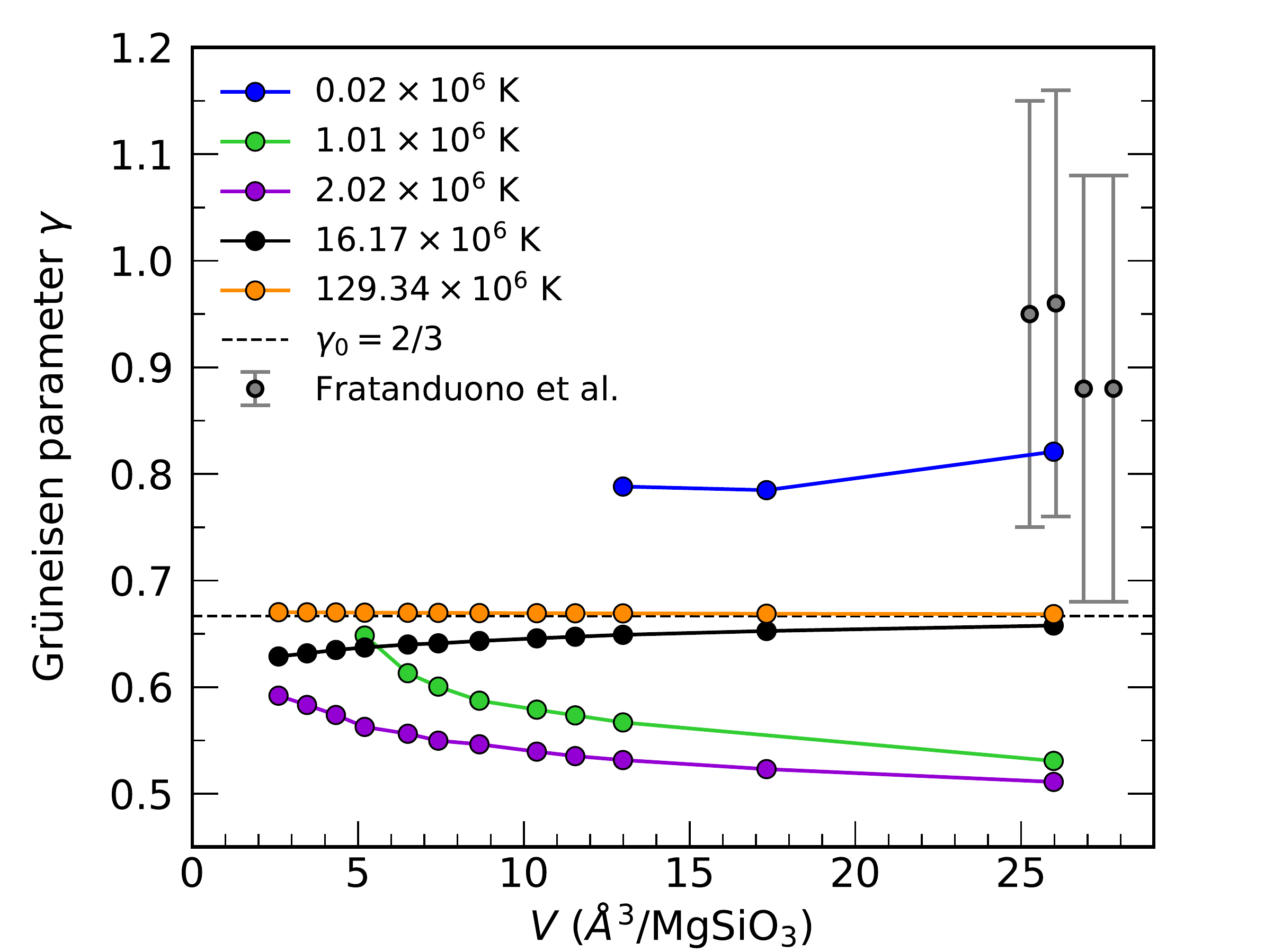}
\caption{
Gr\"uneisen parameter of MgSiO$_3$ as a function of volume for different
temperatures.  The horizontal, dashed line represents high temperature limit
(the Gr\"uneisen parameter of the ideal gas), $\gamma_0=2/3$.  Values of
Gr\"uneisen parameter along the principal Hugoniot from shock compression
experiments~\cite{Fratanduono2018}  (shown in grey circles with error bars)
correspond to pressures between 230--380 GPa and temperatures between
6200--10000 K.
}
\label{fig:Gruneisen}
\end{figure}

In Fig.~\ref{fig:Gruneisen} we show the Gr\"uneisen parameter,
calculated from our EOS using Eq.~\eqref{eq:Gruneisen},
as a function of volume at different temperatures.
First principles simulations and experiments report that in liquid MgSiO$_3$,
contrary to the usual trend in solids, $\gamma$
increases upon compression~\cite{Mosenfelder2009,DeKoker2009} for temperatures
up to 8000 K and volumes from 25.8--64.6 \AA$^3$/f.u.  We also observe this
behavior in most of the temperatures analyzed in our study, as shown in
Fig.~\ref{fig:Gruneisen}, with the exception of $2\times10^4$ K, where we
observe that $\gamma$ decreases with upon compression.  However, this behavior
changes dramatically at higher temperatures.  At $5\times10^4$ K,  $\gamma$ is
almost independent of volume, and above $7\times10^4$ K, it increases upon
compression, as it was observed in experiments at much lower densities and
temperatures.
Our results indicate that the Gr\"uneisen parameter can decrease along an isentrope
for temperatures below $3\times10^4$ K, but increases along the isentropes
with temperatures between $4\times10^4$ and $4\times10^5$ K.


At even higher temperatures the dependence on volume becomes weaker, and at
$16\times10^6$ K, $\gamma$ decreases upon compression.  If the temperature is
high enough to ionize the K shell of the atoms, as we will disccuss in the
next section, the plasma behaves similar to a gas of free particles, described
by the equation of state $E=\frac32PV$ and the Gr\"uneisen parameter
$\gamma_0=2/3$, which is independent of both temperature and density.

\section{K shell Ionization}\label{sec:Ionization}

\begin{figure*}[!hbt]
\includegraphics[width=15cm]{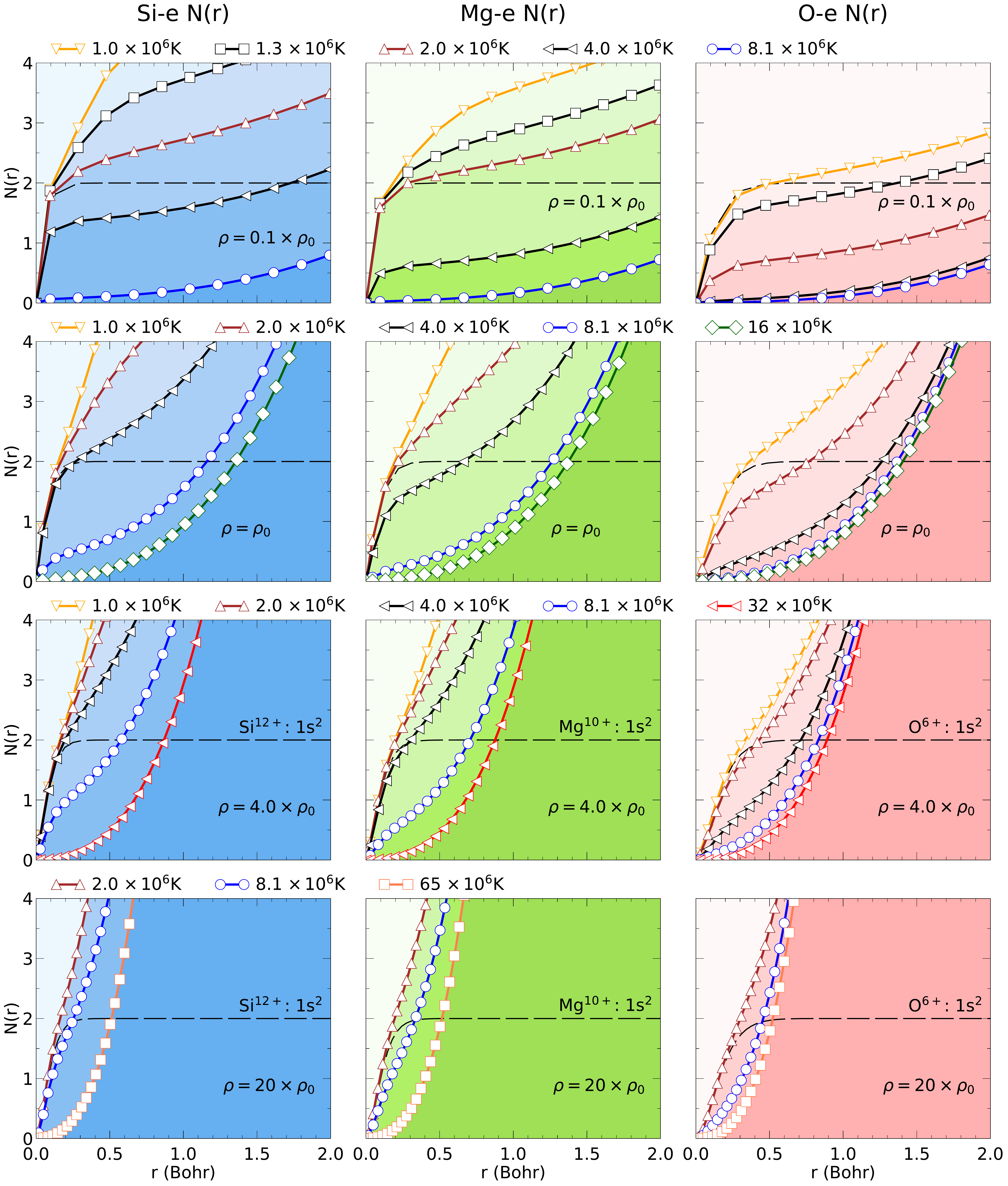}
\caption{
  Integrated nuclear-electron pair correlation, functions $N(r)$,
  computed with PIMC simulations. The three columns correspond to the 
  Mg, Si and O nuclei while the four rows show results for the densities of 0.321
  (0.1-fold compression), 3.21 (1.0-fold compression), 12.83 (4.0-fold
  compression), and 64.16 g cm$^{-3}$ (20-fold compression). For given density, 
  results are shown for the same set of temperatures for all three nuclei. 
  However, these temperatures adjusted with increasing density because the degree of ionization is density dependent. 
  The $N(r)$ represent the average of number of
  electrons contained within a sphere of radius, $r$, around a given nucleus. For comparison we show the corresponding
  functions with thin dashed lines for isolated nuclei with double occupied 1s core states
  that we computed with the GAMESS software~\cite{GAMESS}.
  \label{fig:N(r)}
}
\end{figure*}

From PIMC simulations, a measure of the degree of ionization can be
obtained from the integrated nucleus-electron pair correlation
function, $N(r)$, given by
\begin{equation}\label{eq:N(r)}
N(r)= \left<\frac{1}{N_I}\sum_{e,I}\Theta(r -\|\vec r_e-\vec r_I \|)\right>,
\end{equation}
where $N(r)$ represents the average number of electrons within a sphere of
radius $r$ around a given nucleus of atom of type $I$.  The summation includes
all electron-nucleus pairs and $\Theta$ represents the Heaviside function.
Fig.~\ref{fig:N(r)} shows the integrated nucleus-electron pair correlation
function for temperatures from $1\times 10^6$ K to $65\times 10^6$ K and
densities from 0.321 g$\,$cm$^{-3}$ (0.1-fold) to 64.20 g$\,$cm$^{-3}$ (20-fold
compression). For comparison, the $N(r)$ functions of an isolated nucleus with a
doubly occupied 1s orbital were included. Unless the 1s state
is ionized, its contribution will dominate the $N(r)$ function at small
radii of $ r < 0.2 $ Bohr radii. For larger radii, contributions from other
electronic shells and neighboring nuclei will enter. Still, this is the most
direct approach available to compare the degree of 1s ionization of the three
nuclei. 

At 0.1-fold compression, the comparison with the corresponding curves for the
isolated nuclei shows that the ionization of the 1s states of the Si and Mg
nuclei occurs over the temperature interval from 2.0 to $4.0\times 10^6$ K.
Conversely, the ionization of 1s state of the oxygen nuclei starts already at
1.0 $\times 10^6$ K, which reflects the difference in binding energy that
scales with the square of the nuclear charge, $Z$. Consistent with this
interpretation, one finds that for 4.0 $\times 10^6$ K the Mg nuclei are
slightly more ionized than the heavier Si nuclei, while the ionization of the
oxygen nuclei is essentially complete at this temperature.

When the density is increased from 0.1- to 1.0-fold compression
(second row of panels in Fig.~\ref{fig:N(r)}), the degree of 1s ionization is
reduced. For all three nuclei, the $N(r)$ functions at small $r$ are
closer to doubly occupied 1s state than they were before. This trend
continues as we increase the density to 4.0 and 20-fold
compression. The degree of 1s ionization is consistently reduced
with increasing density
when the results are compared for the same temperature. Most notably we
find the silicon 1s state to be almost completely ionized at 0.1-fold
compression and 8.1 $\times 10^6$ K while very little ionization is
observed at this temperature for 20-fold compression. Similarly,  we
find almost no ionization of the oxygen 1s state at 20-fold
compression and $2.0\times 10^6$ K, while this state is significanly
ionized for 0.1-fold compression at the same temperature. 

For temperatures higher than $32\times10^6$ K, thermal excitations are enough
to fully ionize all atomic species at any of the densities explored, and the
electrons become unbound free particles.  This picture is consistent with our
Gr\"uneisen parameter calculations, which show (see Fig.~\ref{fig:Gruneisen})
that the system has already reached the limiting value of $\gamma_0=2/3$ at
this temperature, consistent with the ideal gas.
We consistently find the degree of 1s ionization to decrease as we
lower the temperature or increase the density in our PIMC
simulations. The trend with density can be interpreted as an
entropy-driven 1s ionization, that can be described by Saha ionization
equilibrium~\cite{Ebeling1976}. With decreasing density, more
free-particle states become available and thus ionization equilibrium
shifts towards higher ionization.

One would expect to find the opposite trend at very high density, where Pauli
exclusion effects cause the 1s state energy to rise, generating a higher degree
of 1s ionization.  However, in our simulations the density is not sufficiently
high for the 1s states of the different nuclei to significantly overlap and
cause ionization by this mechanism.  These results are compatible with the
ionization profile of pure oxygen~\cite{Driver2015b}, where no pressure
ionization of the K shell at $1\times10^6$ K was observed in a similar range of
densities. This analysis does not rule out the possibility of pressure
ionization to occur for higher-energy, more delocalized electronic states.

We find no significant ionization of the 1s orbital at $T=10^6$ K at
any of the densities explored, which indicates that these inner
electrons do not contribute to the thermodynamic properties of the
system at this temperature. DFT pseudopotentials with a  helium core
should, therefore, be sufficient to represent MgSiO$_3$ at these
conditions accurately.  This is not the case for lighter materials,
such as B and LiF~\cite{Zhang2018,Driver2017}, where a temperature
$10^6$ K is enough to cause partial ionization of the K shell due to
the smaller number of electrons that those ions have.  Heavier
elements such as aluminium, to the contrary, require temperatures above $4\times 10^6$
K to ionize the 1s electrons.

\section{Pair correlation functions}

The radial pair correlation function $g_{\alpha\beta}(r)$ is a measure of the atomic coordination,
which depends on temperature and density. It can be interpreted as the probability
of finding an particle of type $\alpha$ at distance $r$ from a particle of type $\beta$.
The nuclear pair-correlation function is given by
\begin{equation}
g_{\alpha\beta}(r)= \frac{V}{{4\pi r^2N_\alpha N_\beta}}\left<\sum_{i=1}^{N_\alpha}\sum_{j\neq i}^{N_\beta}\delta\left(r-\|\vec r_{ij}\|\right)\right>,
\end{equation}
where $N_\alpha$ and $N_\beta$ are the total number of nuclei of type $\alpha$ and $\beta$,
respectively, $V$ is the cell volume, and  $\vec r_{ij}=\vec r_i-\vec r_j$
the separation between atoms $i$ and $j$.

\begin{figure}[!hbt]
\includegraphics[width=9cm]{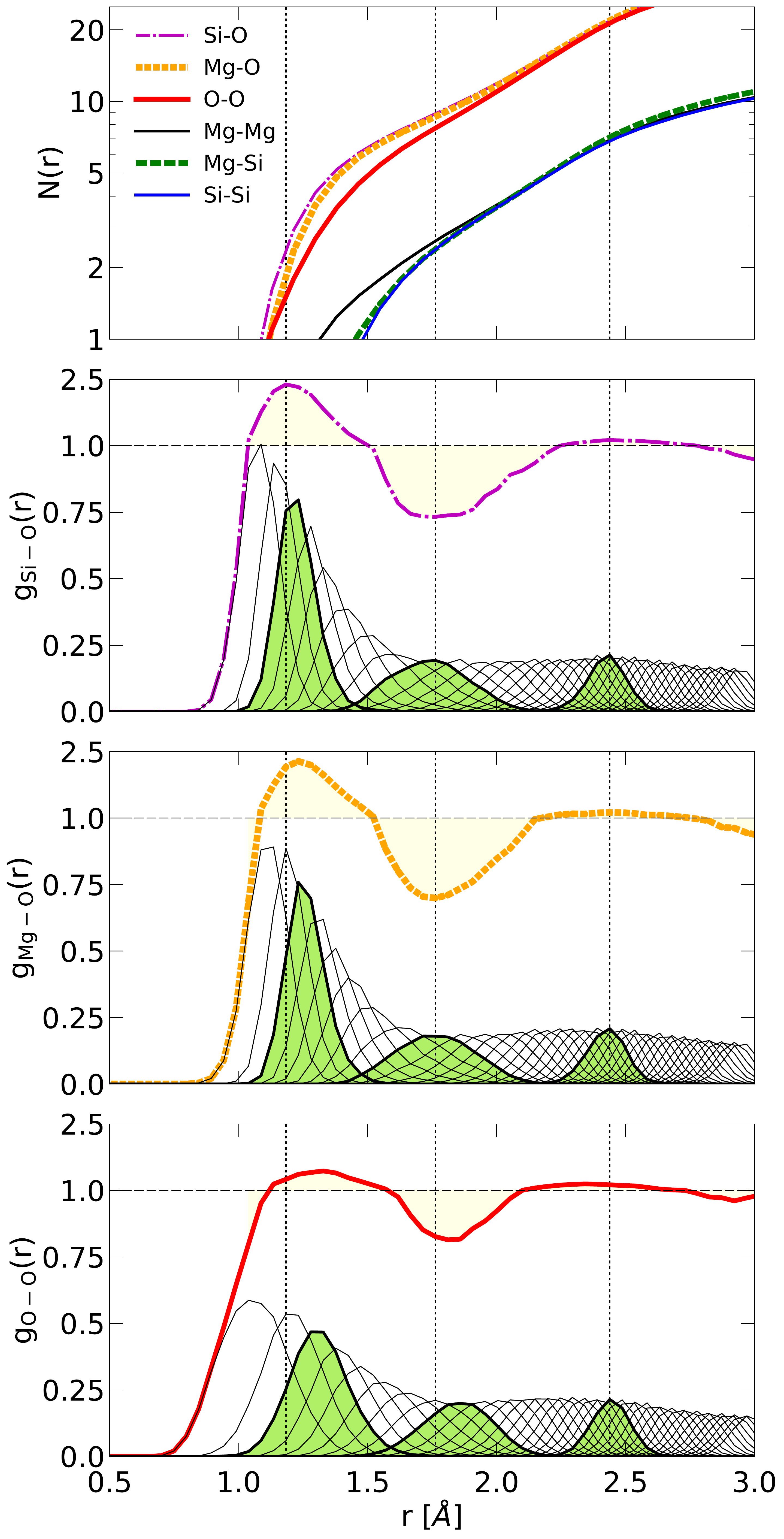}
\caption{$N(r)$ and $g(r)$ correlation functions for liquid MgSiO$_3$
  at $T$ = 50$\,$523 K and $\rho$=19.247 g$\,$ cm$^{-3}$ (6-fold
  compression). In the three lower panels, the $g(r)$ functions were
  split into the contributions from the $n$th nearest neighbors. The
  functions of the 3rd, 9th and 21st neighbors were shaded because
  their locations respectively correspond to first maximum, first
  minimum, and second maximum of the Si-O $g(r)$ function. Their
  locations are also marked by the vertical dotted lines. To improve
  the readability, the scales of the Y axes in the three lower panels
  were split into two separate linear parts, one from 0 and 1, and a
  compressed region from 1 and 2.5.
}\label{gorSplit}
\end{figure}

In Fig.~\ref{gorSplit}, we compare the $N(r)$ and $g(r)$ functions
that we derived with DFT-MD simulations for one $T$-$\rho$ point. The
purpose of this comparison is to analyze how many nearest neighbors
contribute to the various shells of neighboring atoms that appear as
maxima in the pair correlation functions. The $N(r)$ function can be
derived by applying Eq.~\ref{eq:N(r)} to different pairs of nuclei. When
$r$ is set to the first $g(r)$ minimum, the value of $N(r)$ is
commonly referred to as coordination number. The six $N(r)$ functions in
Fig.~\ref{gorSplit} are split into two groups. Functions that involve
oxygen nuclei are much higher because there are three times as many nuclei
that contribute. The Si-O $N(r)$ function rises most quickly with
increasing $r$, reaching a value of $N$=2.44 neighbors for $r=1.18$~\AA,
where the corresponding $g(r)$ function reaches its first maximum of
2.38. This is the most positive nuclear correlation in this dense, hot
fluid. It still carries a signature of the strong Si-O attraction that
leads to the formation of rigid SiO$_4$ tetrahedra that dominate the
coordination in MgSiO$_3$ liquids and solids at much lower temperature
and pressure~\cite{Stixrude2005}. Nevertheless MgSiO$_3$ liquid is
much more disordered at the extreme conditions that we consider in
this article. If one splits the $g(r)$ function into the contributions
from the $n$th nearest neighbor, one finds that the first maximum of
total $g(r)$ functions at $r=1.18$~\AA~falls in between the peaks of
the contributions from the second and third neighbors, as one would
have expected for a value of $N$=2.44. Less expected was how much
overlap there is between contributions from various neigbors. At
$r=1.18$~\AA~there are contributions from up to five oxygen
atoms. Similar if one goes out to the first $g(r)$ minimum at
$r=1.76$~\AA, one finds that contributions from the 9th neighbor dominate
but contributions from the 6th through 12th are still relevant.
At the second $g(r)$ maxium, located
at $r=2.44$~\AA, contributions from the 21st
neighbor dominate.

As expected, we also find a positive correlation between Mg and O
nuclei but it is not quite as strong as that between Si and O
nuclei. The first $g_{\rm Mg-O}(r)=2.28$ maximum occurs at slightly
larger distance of $r=1.23$~\AA. It falls again in between peaks of
the contributions from the second and third neighbors.

The oxygen-oxygen pair correlation function is a bit different but
still positive. Its first maximum is much lower, $g_{\rm O-O}(r)= 1.44$, and occurs only
at large separations of $r=1.33$~\AA. It coincides with the peak in
the $g(r)$ contribution function from the third neighbor.

\begin{figure*}[!hbt]
\includegraphics[width=9cm]{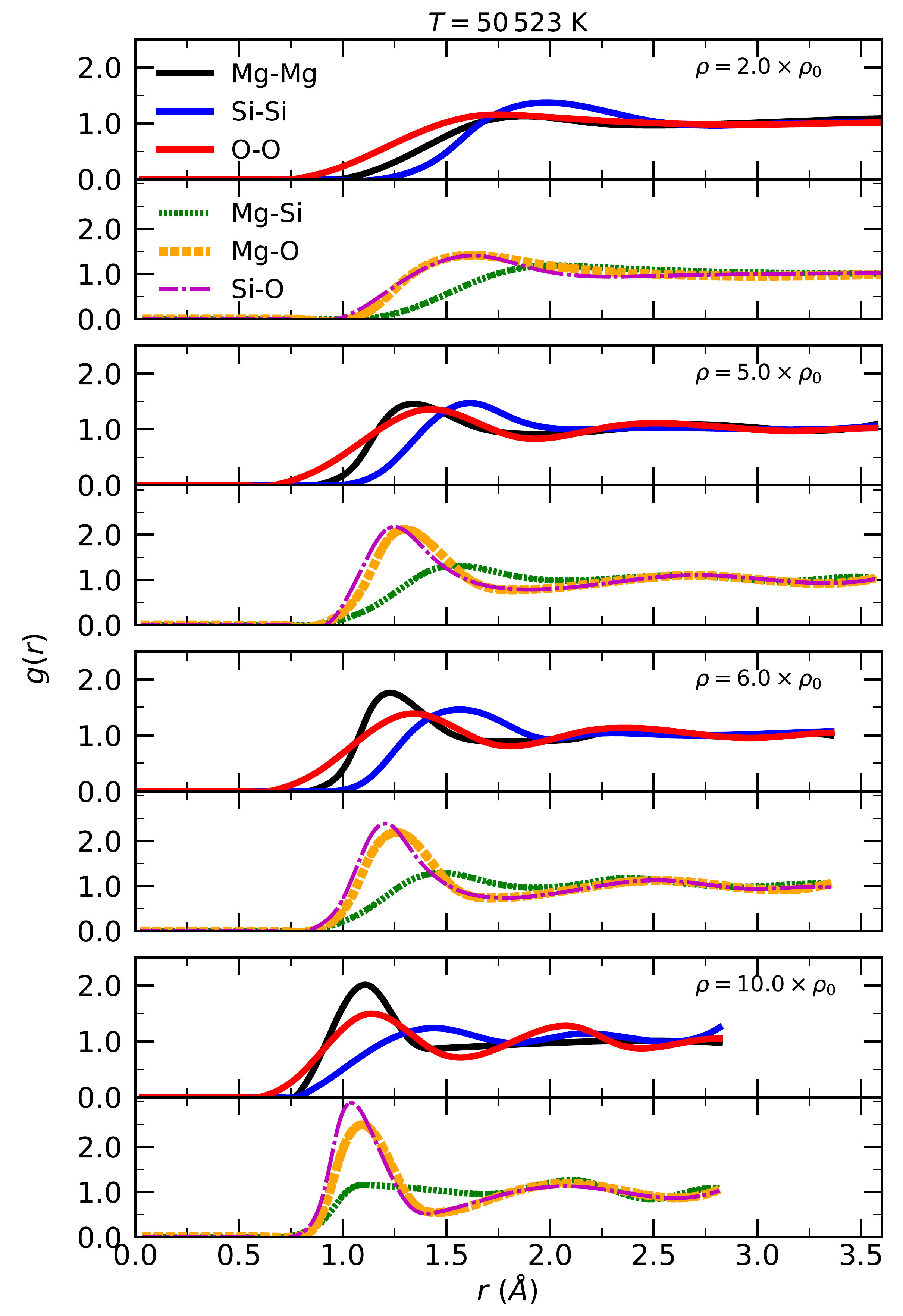}
\includegraphics[width=9cm]{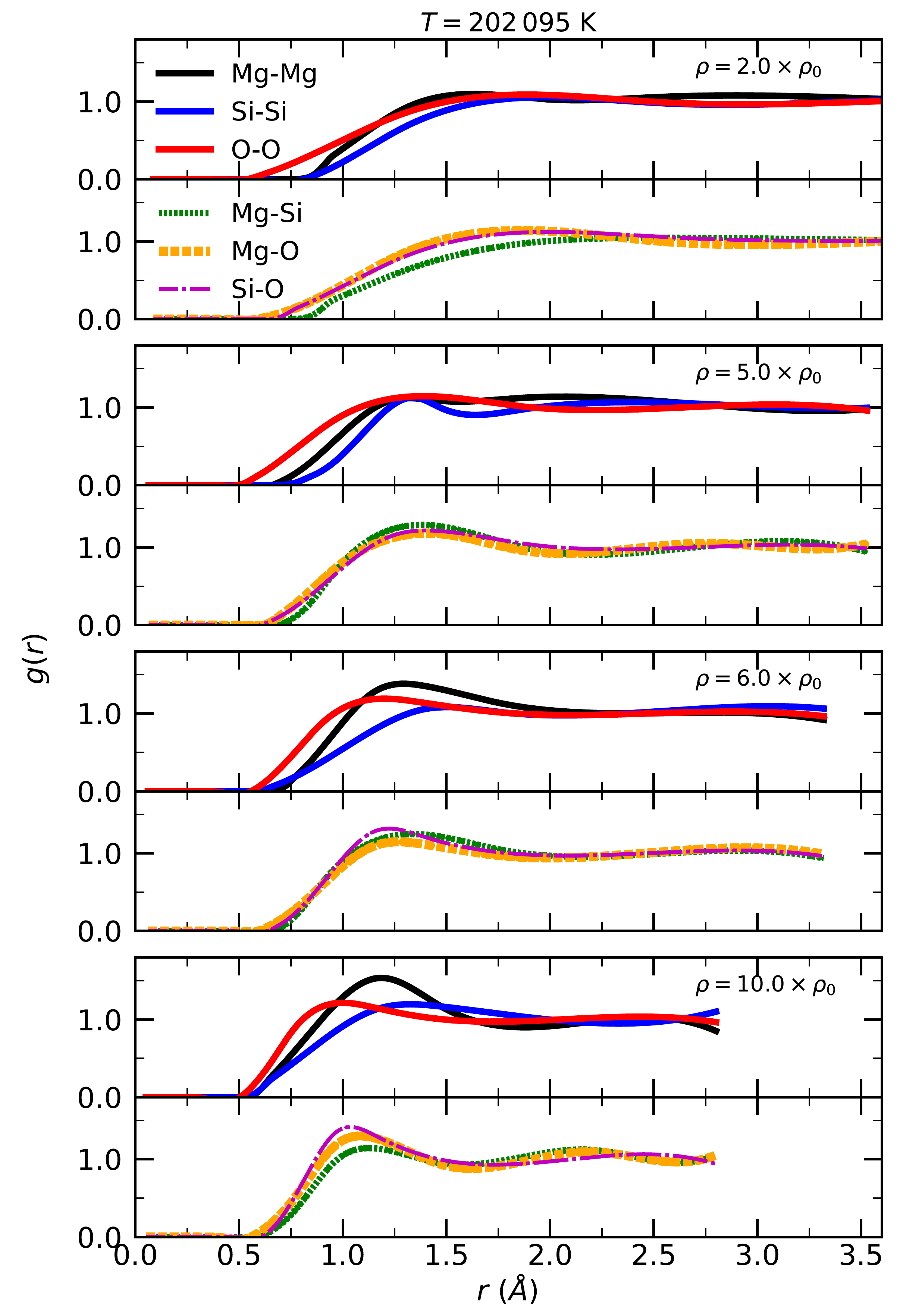}
\caption{
Nuclear radial distribution functions computed with DFT-MD simulations of liquid MgSiO$_3$
at a fixed temperatures of $50\,523$ K and $202\,095$ K.
Functions are calculated in 65-atom cells and compared for densities of (from top to bottom)
6.41 (2-fold, top), 16.04 (5-fold), 19.25 (6-fold), and 32.08 g cm$^{-3}$ (10-fold).
}\label{gor}
\end{figure*}

In figure~\ref{gor}, we compare the nuclear pair correlation function
for 2-, 5-, 6- and 10-fold compression and two temperatures of
$50\times10^3$ and $T=202\times10^3$ K.  At 2-fold compression, the
$g(r)$ function shows the profile of a typical liquid.  The Mg-O and
Si-O bond lengths are approximately equal as the location of
respective first peaks indicate.  As density increases, the atoms get
closer together and these two peaks shift, leading to a stronger
shortening for the Si-O bond than for Mg-O bond. For the other pairs
of species, the first peak of the radial distribution function
localizes at smaller distances, becoming more pronounced as the
density increases. This is evidence for stronger correlations at high
density. This trend is also seen for the Mg-Mg, Si-Si, and O-O pair
correlation functions.

When the temperature is increased from $50\times10^3$ K to
$202\times10^3$ K, the correlation effects are reduced. Most notably
one finds that the Mg-Si, Mg-O, and Si-O pair correlation functions
become fairly similar to each other, while they were rather different
at $50\times10^3$ K.

\section{Shock Hugoniot Curves}
Dynamic shock compression experiments are a direct way to determine
the equation of state of hot, dense fluids by only measuring the shock
and particle velocities. Such experiments are often used to determine
the principal Hugoniot curve, which is the locus of all final states
that can be obtained from different shock velocities~\cite{Fratanduono2018}.

Initially, the sample material has the internal energy, pressure, and
volume, $\{E_0,P_0,V_0\}$. Under shock compression, the material
changes to a final state denoted by $\{E(\rho,T),P(\rho,T),V\}$.  The
conservation of mass, momentum, and energy across the shock front leads
to the Rankine-Hugoniot relation~\cite{Ze66},

\begin{equation}
\left[E(\rho,T)-E_0\right] + \frac{1}{2} \left[P(\rho,T)+P_0\right]\left[V-V_0\right] = 0.
\label{hug}
\end{equation}

\begin{figure}[!hbt]
\includegraphics[width=8.5cm]{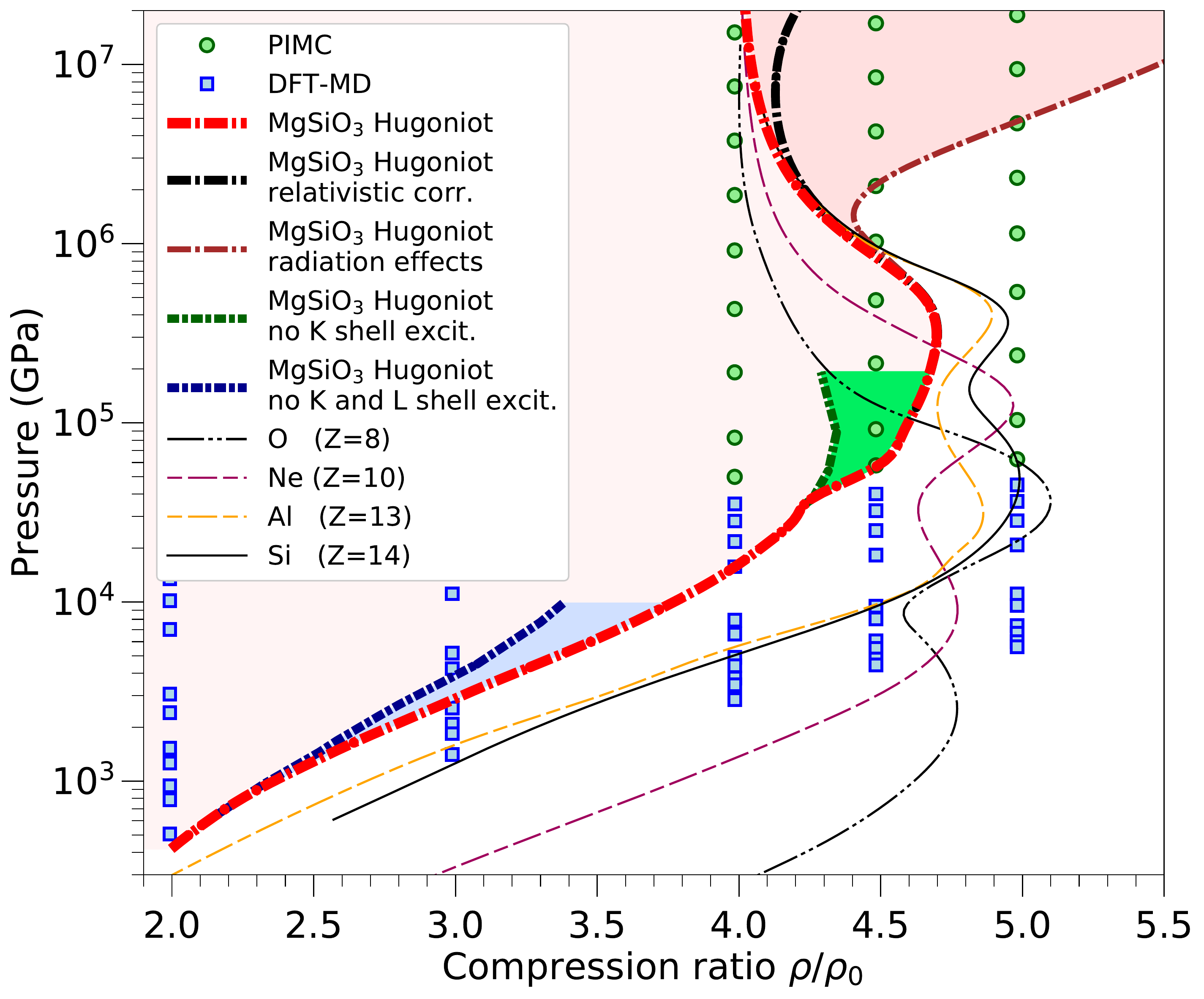}
\caption{ Comparison of the principal shock Hugoniot curves of
  MgSiO$_3$ with those of carbon~\cite{Driver2012},
  oxygen~\cite{Driver2015b}, neon~\cite{Driver2015},
  aluminum~\cite{Driver2018}, and
  silicon~\cite{MilitzerDriver2015,Driver2017b}.  The
    two dash-dotted curves and shaded regions show Hugoniot curves
    without any electronic excitations and just without K shell
    ionization. Electronic excitations increase the shock compression. 
   Without them, a maximum compression ratio of 4.7 could not be attained.
\label{fig:Pvsratio}}
\end{figure}

\begin{figure}[!hbt]
\includegraphics[width=8.5cm]{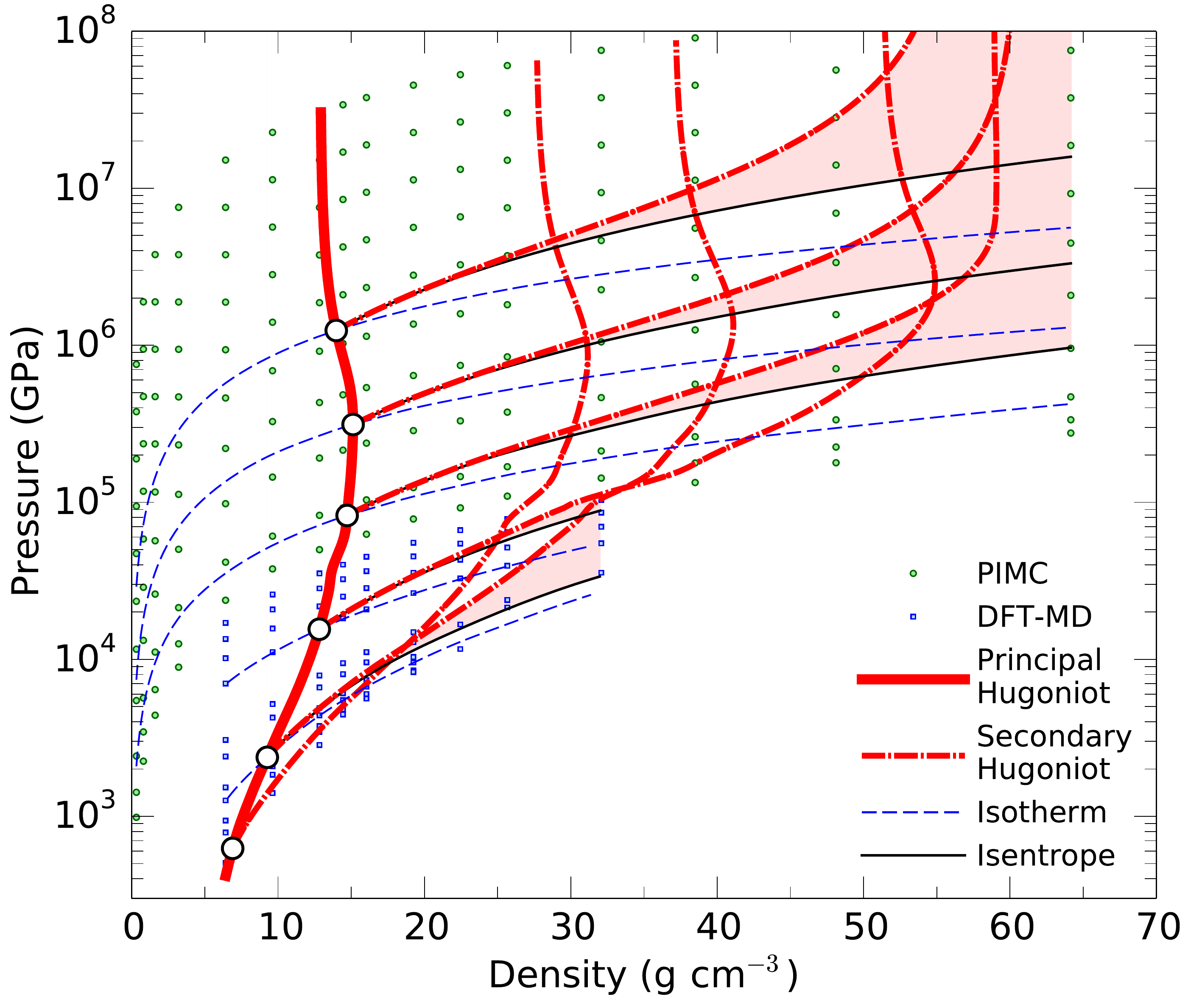}
\caption{ Comparison of principal and secondary shock Hugoniot curves 
with isotherms and isentropes in pressure-density space.
\label{fig:Pvsrho}}
\end{figure}

Here, we solve this equation using our computed EOS table that we provide as
Supplemental Material~\cite{SuppMat}. We obtain a continuous Hugoniot curve by
interpolating $E(\rho,T)$ and $P(\rho,T)$ with 2D spline functions of $\rho$
and $T$. We have compared several different interpolation algorithms and find
the differences are negligible because our EOS table is reasonably dense.  For
the principal Hugoniot curve of solid enstatite, we used $P_0=0$, the ambient
density $\rho_0=3.207911$ g$\,$cm$^{-3}$ ($V_0=51.965073$ \AA$^3$/f.u.), and
initial internal energy $E_0$ = -35.914 eV/f.u. + $\Delta
E$~\cite{Militzer2013d}, where $\Delta E$ is the shift applied to DFT-MD
energies defined in section~\ref{sec:EOS}. The resulting Hugoniot curve has
been added to Figs.~\ref{fig:Tvsrho}, \ref{fig:PvsT}, \ref{fig:Pvsratio}, and
\ref{fig:Pvsrho}.

The principal Hugoniot curve in Fig.~\ref{fig:Pvsratio} exhibits a wide
pressure interval where the compression ratio exceeds 4.0, the value
for an ideal gas. Such high compression values are the result of
excitations of internal degrees of freedom~\cite{Mi06,Mi09}, which
increase the internal energy term in Eq.~\eqref{hug}. Consequently, the
second term in this equation becomes more negative, which reduces the
volume $V$ and thus increases the compression ratio. At a pressure of
15956 GPa and a temperature of 512$\,$000$\,$K, the shock compression
ratio starts to exceed 4, which are conditions where the $L$ shell
electrons are ionized. The bulk of the high compression region is
dominated by the ionization of the K shell (1s) electrons of the three
nuclei. We see one broad region of increased compression instead of
three separate peaks, one for each nucleus. We conclude that the
ionization peaks are merged.

The highest compression ratio of 4.70 is reached for 5.14$\times 10^7$
K and 299$\,$000 GPa, which coincides in pressure with the upper
compression maximum of the shock Hugoniot curve of pure silicon, which
has also been attributed to K shell
ionization~\cite{MilitzerDriver2015}. Based on this comparison and the
K shell ionization analysis of MgSiO$_3$ in Fig.~\ref{fig:N(r)} we
conclude that the upper part of the high compression region in
Fig.~\ref{fig:Pvsratio} is dominated by the ionization of the K shell
electrons of the Si and also the Mg nuclei, because their $N(r)$
curves in Fig.~\ref{fig:N(r)} are fairly similar. The lower end of the
broad compression peak in Fig.~\ref{fig:Pvsratio}, around
6$\times$10$^4$ GPa and 1.4$\times$10$^6$ K, marks the beginning of
the K shell ionization of the oxygen ions as Fig.~\ref{fig:N(r)}
confirms. However, in shock compressed pure oxygen, the K shell
ionization peak occurs for lower P and T. We attribute this difference
to interaction effects in hot, dense MgSiO$_3$ that can shift the
compression peaks along the Hugoniot curve to higher temperatures and
pressures and reduce the peak compression~\cite{Mi06,Mi09}. It should
also be noted that the highest compression ratio of 4.70 is reached
when the K shell electrons of the Si and Mg nuclei are ionized, not
for the lower temperature at which the K shell electrons of oxygen are ionized,
even though three out of five nuclei are of that type and one could
have predicted that their ionization leads to the largest compression. 

We performed additional DFT-MD calculations without any electronic
excitations in order to determine their effect on the shock Hugoniot
curve. In Ref.~\cite{Mi06}, it was shown that
electronic excitations increase to shock compression ratio of helium
to 5.24 while the shock Hugoniot curve without electronic excitations
never exceed 4-fold compression. In Fig.~\ref{fig:Pvsratio}, we show a very similar
behavior for shock compressed MgSiO$_3$. Electronic excitations start to
matter at approximately 30,000 K, 2.3-fold compression, and 850
GPa. With increasing temperature, electronic excitations become more
importance and the gap between the Hugoniot curves with and without
excitations widens. At $10^6$ K, a shock Hugoniot curve without
excitations would yield a pressure of 7700 GPa, $u_p=40.9$ km/s,
$u_s=58.7$ km/s and compression ratio of only 3.3 while with
excitations, the compression ratio is 4.3 and thus the pressure
reaches a much higher value of 38$\,$000 GPa while the particle 
and shock velocities attain much higher values of $u_p=95$ km/s, $u_s=124$
km/s. These differences are a bit smaller if one compares the
predictions for given particle velocity of $u_p=40.9$ km/s, rather
than for constant temperature. With electronic excitations, an
increased compression ratio of 3.6 is predicted while one obtains
slightly reduced values of pressure (7420 GPa) and the shock speed
($u_s=56.6$ km/s). However, the temperature is much lower (280$\,$000
K) than is predicted without excitation ($10^6$ K). This underlines that
electronic excitation significantly affect the final state in shock
compression experiments of dense silicates.

In Fig.~\ref{fig:Pvsratio}, we also show a shock Hugoniot curve that includes L shell
but no K shell ionization. This curve was derived from VASP DFT-MD
simulations that relied on pseudopotentials with a frozen K shell
electrons. At 4.26-fold compression, 37700 GPa, and 1.01 $\times 10^6$ K,
this curve starts to deviate from our original Hugoniot curve that
included the K shell ionization. It is the ionization of this shell
that introduces a shoulder into the Hugoniot curve and increases the
compression to a maximum value of 4.7.

Very approximately, we added relativistic and radiation effects to the
Hugoniot curves in Fig.~\ref{fig:Pvsratio}. Under the assumption of
complete ionization, the relativistic corrections were derived for an
ideal gas of electrons. This increases the shock compression ratio for
$P > 4 \times 10^6$ GPa and $T > 7 \times 10^7$ K.
Considering an ideal black
body scenario, we derived the photon contribution to the EOS using
$P_\text{radiation}=(4\sigma/3c)T^4$ and
$E_\text{radiation}=3 P_\text{radiation}V$, where $\sigma$ is the
Stefan-Boltzman constant and $c$ is the speed of light in vacuum.
We find that radiation effects are important only for temperatures
above $3\times10^7$ K, which are well above the temperature necessary to
completely ionize the 1s orbitals of all atomic species.

In Fig.~\ref{fig:PvsT}, we can observe how our calculated Hugoniot overlaps
with the experimental data from Fratanduono et al.~\cite{Fratanduono2018} who
performed laser-driven shock experiments on enstatite to obtain a continuous
measurement of the principal Hugoniot curve.  In these experiments, enstatite
was shocked up to 600 GPa reaching temperatures as high as $2\times10^4$ K.
Using a Gr\"uneisen parameter model coupled to an EOS, Fratanduono et al.
derive isentropic profiles for liquid MgSiO$_3$ close to the melting curve.
Their findings show that the melting curve and the isentropic temperature
profiles, shown in Fig.~\ref{fig:PvsT}, are shallower than
previous DFT-MD predictions~\cite{Stixrude2014} and nearly parallel to each
other, which can have substantial implications for the interior of rocky
exoplanets, such as the possible crystallization of a deep silicate
mantle over a wide range of temperatures.

To provide a guide for future ramp compression experiments, we also plot
different isentropes, derived from the relationship
$\left.\frac{dT}{dV}\right|_S=-T\left.\frac{dP}{dT}\right|_V/\left.\frac{dE}{dT}\right|_V$
and added them to Figs.~\ref{fig:Tvsrho}, \ref{fig:PvsT}, and \ref{fig:Pvsrho}.
We find that the slope of the isentropes does not strongly depend on
temperature, even though we compare conditions with differing degrees of
ionization. Our results imply that the temperature rise with pressure along the
isentropes approximately follows a power law, $T\propto P^\alpha$, with an
exponent $\alpha=0.309$ below $10^6$ K, increasing only up to $\alpha=0.399$
for temperatures above $10^7$ K.  This provides simple rule for obtaining
isentropic profiles in MgSiO$_3$ with wide-range validity, without the need of
relying in approximate models.

In Fig.~\ref{fig:Pvsrho}, we show a number of double-shock Hugoniot
curves. Various points on the principal Hugoniot curve were chosen
as initial conditions for a second shock that compresses the material
again, reaching densities that are much higher than those that can probed with
single shocks. If one starts from the high compression point on the
principal Hugoniot curve, one can reach densities of 60
g$\,$cm$^{-3}$. However, the compression ratio is typically not as
high because the strength of the interaction effects increases and
this lowers the compression ratio. For the secondary shock Hugoniot
curves that we show in Fig. ~\ref{fig:Pvsrho}, the maximum compression
ratio varied between 4.44 and 4.01 while the maximum compression ratio
of the principlal Hugoniot curve was 4.70.

Fig.~\ref{fig:Pvsrho} also compares our secondary Hugoniot curves
with our isentropes and isotherms. For weak second shocks, the secondary
Hugoniot curves and isentropes almost coincide, which implies that the
second shock produces very little nonreversible heat. As the
strength of the second shock increases, more and more nonreversible
heat is generated.

\begin{figure}[!hbt]
\includegraphics[width=9cm]{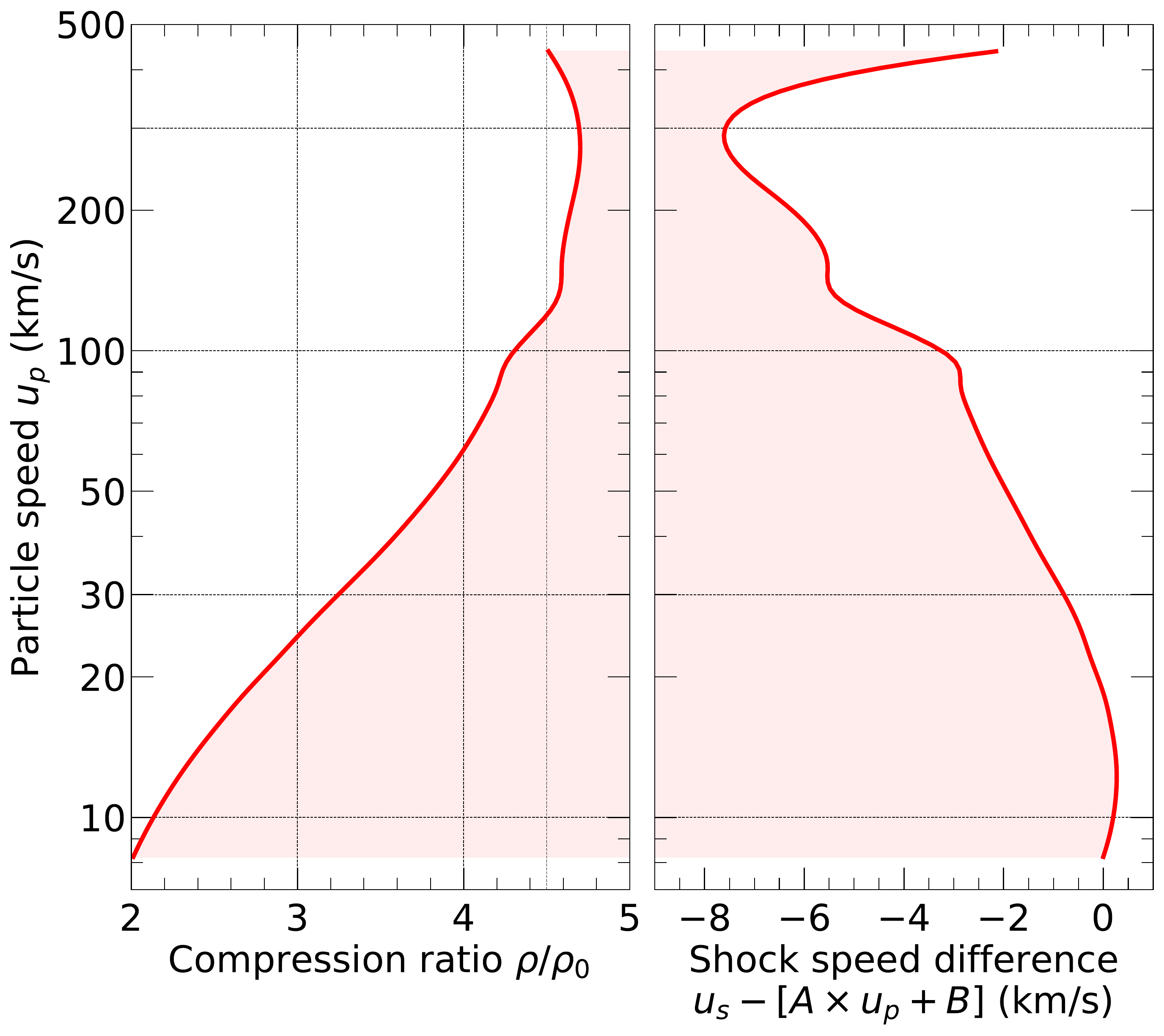}
\caption{On the left, particle velocity $u_p$ is shown a function of compression ratio.
On the right, we show the deviations from a linear $u_p$-$u_s$ fit. The coefficients were $A=1.2764$ and $B=8.2315$ km/s. }\label{fig:UsUp}
\end{figure}

To provide a direct comparison with experiments, we also derived the particle
velocity, $u_p$, and shock velocity, $u_s$, along the principal Hugoniot curve. The left panel of Fig.~\ref{fig:UsUp} shows
the particle velocity as a function of compression ratio, which allows on to related the prediction to Fig.~\ref{fig:Pvsratio}. 
It is often found that $u_p$ and $u_s$ follow an approximately
linear relationship over a wide range of
conditions~\cite{Nellis1984,Fratanduono2018}. However, one does not
expect a linear relationship to hold perfectly when electronic
excitations introduce distinct increases in compression at
well-defined temperature/pressure intervals. Therefore we first fit a
linear $u_p$-$u_s$ relation for our computed Hugoniot curve and then
plot the deviation from it in the right panel of
Fig.~\ref{fig:UsUp}. The comparison of both panels allows us to
correlate deviations from linear $u_p$-$u_s$ relation with changes in
compression. For example the onset of the K shell ionization that
introduces a bump into the Hugoniot curve at 4.3-fold compression also
leads to a bump in $u_s$ for $u_p$ = 90 km/s. Similarly, when the K
shell ionization of the oxygen atom increase the compression ratio to
4.6, we see a reduction in $u_s$ for $u_p$ = 140 km/s. Finally, the
ionization of the Mg and Si K shell electrons that leads to the
compression maximum of 4.7 $\times \rho_0$ also leads to a reduction
in $u_s$ for $u_p$ = 270 km/s. For even higher particle velocities,
the system approaches the states of a completely ionized plasma where the shock compression ratio is gradually reduced to 4 and our linear $u_p$-$u_s$ relation does no longer hold.

\section{Specific Heat}
\begin{figure}[!hbt]
\includegraphics[width=9cm]{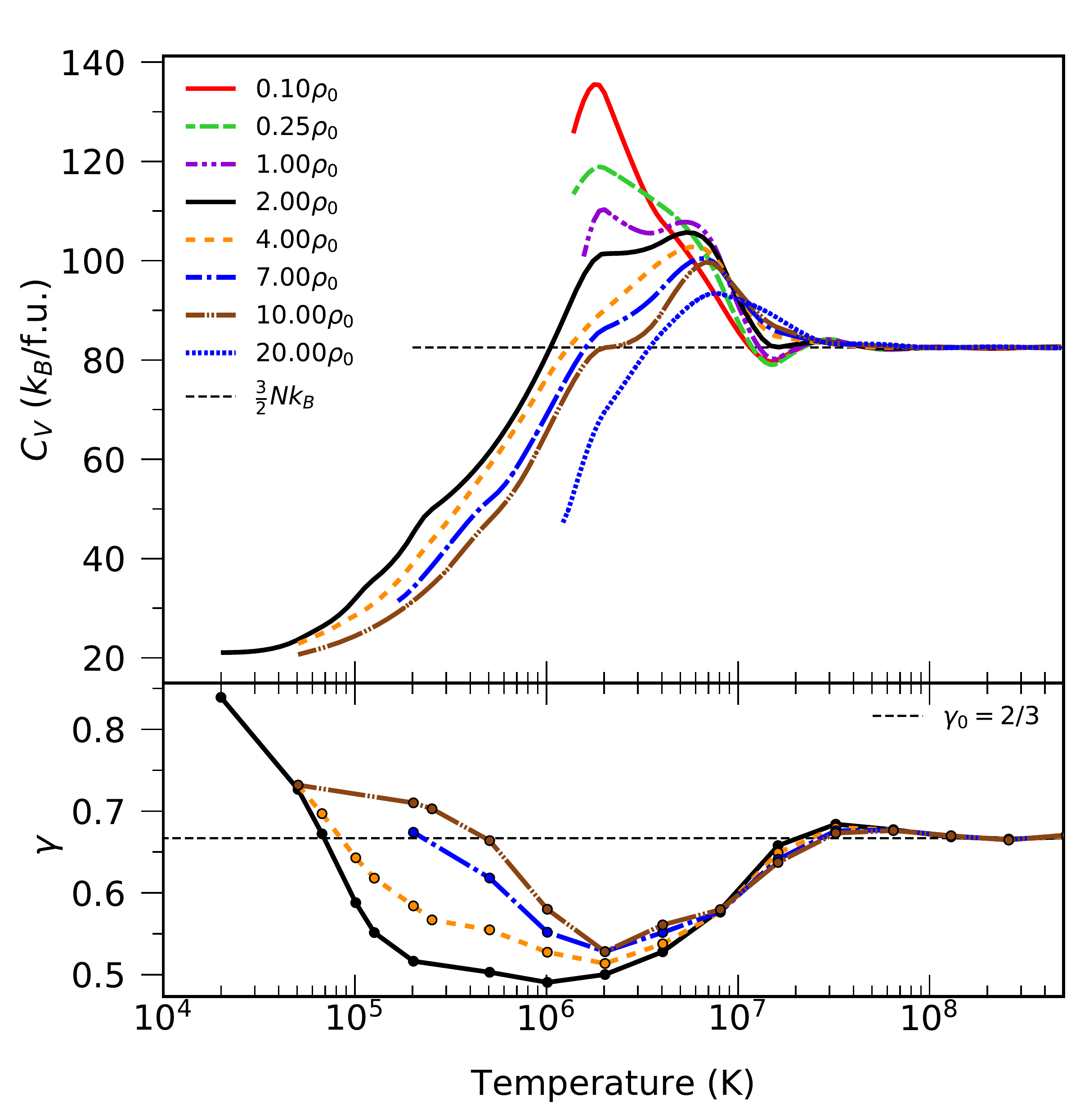}
\caption{Heat capacity and Gr\"uneisen parameter temperature dependence,
derived from DFT-MD and PIMC calculations at various densities. The horizontal
dashed line in the upper panel represents the high temperature limit of
$C_V=\frac32 Nk_B$ where $N=55$ is the total number of free particles per
MgSiO$_3$ formula unit, with 5 ions and 50 electrons.  Values above this line
mark the temperature region of ionization where the intenal energy increases
significantly.  In the lower panel, the Gr\"uneisen parameter of the ideal gas,
$\gamma_0=2/3$, also represents the high temperature limit.
}\label{fig:Cv}
\end{figure}

The specific heat, $C_v=\left.\frac{\partial E}{\partial T}\right|_V$ is shown in Fig.~\ref{fig:Cv}
as a function of temperature for various densities.
For temperatures below 10$^5$ K, our calculations show that
the value of $C_v$ approaches 21 $k_B$/f.u. (4.2 $k_B$/atom) at 2-fold compression
(6.42 g cm$^{-3}$), which is in agreement with previous DFT calculations~\cite{DeKoker2009}
and recent experimental measurements~\cite{Fratanduono2018} along the Hugoniot at similar conditions.
At very high temperatures, where the all atomic species are completely ionized,
we recover the expected nonrelativistic limit of $\frac 32 Nk_b$, where
$N=55$ is the total number of free particles of the MgSiO$_3$ system (5 nuclei and 50 electrons).

As electrons become free with increasing temperature, the specific heat increases,
reaching a local maximum at around $T\approx2\times10^6$ K for the density range of 0.1--2$\rho_0$,
which reflects the ionization of K shell electrons of the oxygen atoms.
This peak dissapears almost completely at $4\rho_0$ because this compression prevents
the oxygen K shell electrons from becoming ionized, as we discussed in the previous section.
A second maximum appears around $T\approx7\times10^6$ K, which can be associated with 
the almost simultaneous ionization of Mg and Si K shell electrons, as we showed in
Fig.~\ref{fig:N(r)}.

In the bottom panel of Fig.~\ref{fig:Cv}, we observe that the Gr\"uneisen parameter,
$\gamma,$
decreases with temperature up to approximately $2\times10^6$ K for all
densities considered, due to the increasing value of $C_v$ (see Eq.~\eqref{eq:Gruneisen}).  Above
$2\times10^6$ K, where ionization of the K shells takes place, $\gamma$
increases with temperature due to the decrease in $C_v$  and tends to the ideal
gas limit of $\gamma_0=2/3$, as we have shown in Fig.~\ref{fig:Gruneisen}.


A discontinuity in the principal Hugoniot of liquid MgSiO$_3$ has been observed
around 15000 K and 500 GPa~\cite{Spaulding2012}, which was interpreted as a
liquid--liquid phase transition that could lead to an unusually large increase
of the specific heat. According to this study, $C_v$ could be as large as 90
$k_B$/f.u. (18 $k_B$/atom) at these conditions, a value that is expected only
at temperatures beyond 10$^6$ K, according to our calculations.  However, this
transition has not been confirmed in previous DFT-MD simulations~\cite{Militzer2013d},
and recent experiments~\cite{Fratanduono2018,Spaulding2017} show no anomalies
in the principal Hugoniot that could support this hypothesis. Therefore, we should
expect $C_V$ to be at most 30 $k_B$/f.u. (6 $k_B$/atom) below $100\,000$ K.

\section{Conclusions}

We have constructed a consistent EOS of MgSiO$_3$ over a wide
temperature-density range using DFT-MD and PIMC that bridges the WDM
and plasma regimes. Our results provide the first
detailed characterization of K shell ionization in a triatomic
material. We quantify the degree of ionization and the contribution
from each atomic species to the thermodynamic properties, which, at
the present time, cannot be inferred from the laboratory experiments.
We predict that the maximum compression ratio for enstatite is 4.7,
which is attained for $5.13\times10^6$ K and $3.01\times10^5$ GPa in
the WDM regime. By performing additional calculations without any
electronic excitations or only without K shell excitations, we are
able to determine the conditions where these excitations start to
increase the shock compression. We show that without electronic
excitations the shock compression ratio of MgSiO$_3$ would not exceed
4.0. Excitations of L shell electrons start increase the shock
compression from 30,000 K, 847 GPa, $\rho/\rho_0=2.28$, $u_p=12.2$ km/s
and $u_s=21.7$ km/s onwards, which is within the reach of current
laboratory experiments. It is also interesting to note that we do not
see a separate L shell ionization peak. We conclude that this shell is
ionized gradually, as it occurs in dense carbon and boron
materials~\cite{ZhangCH2017,ZhangCH2018,Zhang2018,ZhangBN2019}. Excitations
of K shell electrons set in at 1.01 $\times 10^6$ K, 37700 GPa,
$\rho/\rho_0=4.26$, $u_p=94.8$ km/s and $u_s=124$ km/s.

We find good agreement between results from PIMC and DFT-MD
simulations, which provides evidence that the combination of these two
different formulations of quantum mechanics can be used to accurately
describe WDM. The precision of first-principles computer simulations will guide the
design of inertial confinement fusion (ICF) experiments under
conditions where the K and L shell electrons are gradually ionized,
which is challenging to predict accurately with analytical EOS models.

We showed that PIMC and DFT-MD simulations produce consistent EOS data
in the 1--$ 2 \times 10^6$ K temperature range, validating the use of
the fixed-node approximation in PIMC and zero-temperature XC
functionals in DFT-MD for warm dense MgSiO$_3$.  We obtain a shock
Hugoniot curve that is consistent with experiments and includes the K
shell ionization regime of the three atomic species. Their ionization
leads a one broad peak of high compression ratios between 4.5 and
4.7. The maximum compression is reached for higher temperatures, where
the Mg and Si atoms are ionized, even though there are more oxygen
atoms present and their 1s ionization occurs at slightly lower
temperatures.

Subsequently, we analyzed how close a secondary shock Hugoniont curves
can stay to an isentrope, providing a guide for future ramp
compression experiments.
We also showed that the Gr\"uneisen parameter increases upon compression for
most of the temperatures analyzed in this study, and converges to the ideal gas
limit when temperature reaches $\sim 2 \times 10^7$, consistent with a full
K shell ionization of all atomic species.

Finally, we then studied heat capacity and pair-correlation functions
to reveal the evolution of the fluid structure and ionization
behavior. Overall, we demonstrate that PIMC is an predictive tool to
determine the EOS in the WDM regime. We demonstrated that He-core PBE
functional can accurately describe MgSiO$_3$ up to temperatures of
$\sim 10^6$ K. For higher temperature, the ionization of K shell
electrons significantly affect the thermodynamic properties and the
shock Hugoniot curve of MgSiO$_3$ and the frozen-core approximation in
the pseudopotential no longer valid.

\begin{acknowledgments}
  This work was in part supported by the National Science
  Foundation-Department of Energy (DOE) partnership for plasma science
  and engineering (grant DE-SC0016248), by the DOE-National Nuclear
  Security Administration (grant DE-NA0003842), and the University of
  California Laboratory Fees Research Program (grant
  LFR-17-449059).
	We thank the authors of Ref.~\cite{Fratanduono2018} for providing the
	full data set from their experiments.
	F.G.-C. acknowledges support from the CONICYT
  Postdoctoral fellowship (grant 74160058). Computational support was
  provided by the Blue Waters sustained-petascale computing project
  (NSF ACI 1640776) and the National Energy Research Scientific
  Computing Center.
\end{acknowledgments}

\section*{References}


%

\end{document}